\documentclass[prd,preprint,superscriptaddress]{revtex4}
\usepackage{revsymb}
\usepackage{amsmath}
\usepackage{graphicx}
\usepackage{bm}
\newcommand{\be}{\begin{equation}}
\newcommand{\ee}{\end{equation}}
\newcommand{\ben}{\begin{eqnarray}}
\newcommand{\een}{\end{eqnarray}}

\begin{document}
\title{On holographic dark-energy models}
\author{Sergio del Campo\footnote{E-Mail:
sdelcamp@ucv.cl}} \affiliation{Instituto de F\'{\i}sica,
Pontificia Universidad Cat\'{o}lica de Valpara\'{\i}so, Avenida
Brasil 2950, Casilla 4059, Valpara\'{\i}so, Chile}
\author{J\'{u}lio.C. Fabris\footnote{E-mail: fabris@pq.cnpq.br}}
\affiliation{Universidade Federal do Esp\'{\i}rito Santo,
Departamento
de F\'{\i}sica\\
Av. Fernando Ferrari, 514, Campus de Goiabeiras, CEP 29075-910,
Vit\'oria, Esp\'{\i}rito Santo, Brazil}
\author{Ram\'{o}n Herrera\footnote{E-mail: ramon.herrera@ucv.cl}}
\affiliation{Instituto de F\'{\i}sica, Pontificia Universidad
Cat\'{o}lica de Valpara\'{\i}so, Avenida Brasil 2950, Casilla
4059, Valpara\'{\i}so, Chile}
\author{Winfried Zimdahl\footnote{E-mail: winfried.zimdahl@pq.cnpq.br}}
\affiliation{Universidade Federal do Esp\'{\i}rito Santo,
Departamento
de F\'{\i}sica\\
Av. Fernando Ferrari, 514, Campus de Goiabeiras, CEP 29075-910,
Vit\'oria, Esp\'{\i}rito Santo, Brazil}

\begin{abstract}
Different holographic dark-energy  models are studied from a unifying point of view.
We compare models for which the Hubble scale, the future event horizon
 or a quantity proportional to the Ricci scale are taken as the infrared cutoff length. We demonstrate that the mere definition
 of the holographic dark-energy  density generally implies an interaction with the dark-matter component.
 We discuss the relation between the equation-of-state parameter and the energy density ratio of both components
 for each of the choices, as well as the possibility of non-interacting and scaling solutions.
Parameter estimations for all three cutoff options are performed with the help of a Bayesian statistical analysis, using data from supernovae type Ia and the history of the Hubble parameter.
The $\Lambda$CDM model is the clear winner of the analysis. According to the Bayesian Information Criterion ($BIC$), all holographic models should be considered as ruled out, since the difference $\Delta BIC$ to the corresponding $\Lambda$CDM value is $> 10$.
According to the Akaike Information Criterion ($AIC$), however, we find $\Delta AIC$ $< 2$ for models with Hubble-scale and Ricci-scale cutoffs, indicating, that they may still be competitive. As we show for the example of the Ricci-scale case, also the use of certain priors, reducing the number of free parameters to that of the $\Lambda$CDM model, may result in a competitive holographic model.
\end{abstract}

\maketitle

\section{Introduction}

Observations from Supernova (SN Ia) \cite{01}, large scale
structure \cite{02}, cosmic microwave background \cite{03}, the
integrated Sachs--Wolfe effect \cite{isw}, baryonic acoustic
oscillations \cite{eisenstein} and from gravitational lensing
\cite{weakl} suggest that our current universe is in a state of
accelerated expansion. In most investigations the origin of the
acceleration is attributed  to a mysterious component with a
negative pressure, called dark energy (DE). The preferred
candidate for this entity is a cosmological constant $\Lambda$.
The favored cosmological model is the $\Lambda$-cold-dark-matter
($\Lambda$CDM) model which also plays the role of a reference
model for alternative approaches to the DE problem. Alternative DE
models have been developed since the $\Lambda$CDM model, although
it fits  most observational data rather well, suffers from two
main and lasting shortcomings, namely: the low value of the vacuum
energy and the coincidence problem \cite{1}. In order to address
these two problems, the cosmological constant has been replaced by
a time varying quantity, resulting in dynamical DE models. The
best studied models here are scalar field models which comprehend,
e.g., quintessence \cite{2}, K-essence \cite{3} and tachyon fields
\cite{4}.  In most investigations DM and DE are considered as
independent substances. However, there is no reason to neglect
interactions in the dark sector. Including the possibility of a
coupling between DM and DE is the more general case and gives rise
to a richer dynamics \cite{interaction}. So far, DE and dark
matter (DM) manifest themselves only through their gravitational
action. This circumstance is the motivation for unified models of
the cosmological substratum in which one single component plays
the role of DM and DE simultaneously. A Chaplygin gas represents
the prototype of a unified model \cite{5}, but also
  bulk-viscous models belong to this category (see, e.g., \cite{bv} and references therein).
Among the many different approaches to describe the dark cosmological sector, so called holographic DE models have received considerable attention \cite{6,cohen,li}.
According to the holographic principle, the number of
degrees of freedom in a bounded system should be finite and related to the area of its boundary \cite{7}.
Based on this principle, a field theoretical relation between a short
distance (ultraviolet) cutoff and a long distance (infrared)
cutoff was established \cite{cohen}. This relation ensures that  the energy in a box of size $L$ does not
exceed the energy of a black hole of the same size. Applied to the dynamics of the Universe, $L$ has to be a
cosmological length scale. Different choices of this cutoff scale result in different DE
 models. If one identifies L with the
Hubble radius $H^{-1}$, the resulting DE density, corresponding to the ultra-violet cutoff,
will be close to the observed effective cosmological constant \cite{cohen}.
The possibilities of the particle and the event horizons
as the IR cutoff length were subsequently discussed by Li\cite{li}, who
found that apparently only a future event horizon cutoff can
give a viable DE model. However, afterwards it was recognized,
that a cutoff at the Hubble scale may well result in a realistic cosmological dynamics provided only, that
an interaction in the dark sector is admitted \cite{DW}.
More recently, yet another possibility was proposed, namely a cutoff scale, given by the Ricci scalar curvature
\cite{021,022}, resulting
in  so-called holographic Ricci DE
models.  Holographic DE model have
been tested and constrained by various astronomical observations
\cite{obs}, in some cases also including spatial curvature \cite{cont}.

A special class are models in which  holographic DE is allowed to interact with DM \cite{DW,8,9,10,HDE}.
In this article, we compare three different approaches to (generally interacting) DE models. We study the
relation between the energy density ratio of DM and DE and the equation-of-state (EoS)
parameter for general interactions in each of these cases. Also
solutions for the interaction free limits and scaling solutions
(constant energy density ratio) are obtained as special cases.
We clarify the role of potential interactions in the dark sector. In particular, we point out that any interaction model,
including also the non-interacting limits, introduces relations between the EoS parameter and the matter content, which constrain the admissible dynamics.
All the cases are confronted with observational data from supernovae type Ia and from the history of the Hubble parameter. A statistical analysis is performed on the basis of both the
Akaike Information Criterion ($AIC$) and the
Bayesian Information Criterion ($BIC$).
It turns out, that a cosmological constant remains the favored DE candidate. 

This paper is organized as follows: In Section \ref{basic} we
provide  the basic relations for an arbitrary cutoff scale. In
Section \ref{hubble} we consider the Hubble-scale cutoff. The cutoff
at the future event horizon is studied in Section \ref{event}, including an analysis of
the non-interacting limit and the possibility of scaling solution. The same features are
investigated for the Ricci-scale cutoff in Section \ref{riccicut}.
In Section \ref{statanalysis} we present the results of a Bayesian statistical analysis for each of the cases. Finally, a summary of the paper is given in Section \ref{conclusions}.

\section{Basic equations}
\label{basic}

We assume the cosmic substratum to be described by a pressureless matter component with energy density
$\rho_{m}$ and a holographic dark energy component with energy density $\rho_{H}$.
The  Friedmann equation for the spatially flat case then is
\\
\begin{equation}
3 H^2 = 8\pi\, G (\rho_{m} + \rho_{H})  \  .\label{Fried1}
\end{equation}
Both components are admitted to interact according to
\begin{equation}
\dot{\rho}_{m} + 3 H \rho_{m}  = Q  \label{cons1}
\end{equation}
and
\begin{equation}
\dot{\rho}_{H} + 3 H (1 + w)\rho_{H} = -Q \, . \label{cons2}
\end{equation}
Here we have introduced the equation-of-state (EoS) parameter $w \equiv\frac{p_H}{\rho_{H}}$, where
$p_H$ is the pressure associated with the holographic component.
The total energy density $\rho = \rho_{m} + \rho_{H}$ is conserved.
The Hubble rate changes as
\begin{equation}
\dot{H} = - \frac{3}{2}H^{2}\left(1 + \frac{w}{1+r}\right)  \ , \quad \Rightarrow \quad
\frac{d\ln H}{d\ln a}  = - \frac{3}{2}\left(1 + \frac{w(a)}{1 + r(a)}\right) \ ,
\label{dH}
\end{equation}
where $r=\frac{\rho_{m}}{\rho_{H}}$ is the ratio of the energy densities of both components and
\begin{equation}
\frac{w}{1 + r} \equiv w^{eff}  \label{w}
\end{equation}
is the total effective EoS of the cosmic medium.
The dynamics of the ratio $r$ is determined by
\begin{equation}
\dot{r} = r\,\left(\frac{\dot{\rho}_{m}}{\rho_{m}}-
\frac{\dot{\rho}_{H}}{\rho_{H}}\right). \label{dr}
\end{equation}
Introducing here the balances (\ref{cons1}) and (\ref{cons2}), we arrive at
\begin{equation}
\dot{r} = 3Hr\,\left(1 + r\right)\,\left[\frac{w}{1+r} + \frac{Q}{3H\rho_{m}}\right]\ . \label{dr2}
\end{equation}
Following \cite{cohen,li} we shall write the holographic energy density as
\begin{equation}
\rho_H=\frac{3\,c^2\,M_p^2}{L^{2}} \ .\label{ans}
\end{equation}
The quantity $L$ is the infrared (IR) cutoff scale and
$M_p=1/\sqrt{8\pi\,G}$ is the reduced Planck mass. The numerical constant $c^{2}$ determines the degree
of saturation of the condition
\begin{equation}
L^{3}\, \rho_H\leq M_{Pl}^{2}\, L \, ,    \label{DEineq}
\end{equation}
which is at the heart of any holographic dark energy model and which states that the energy in a box of size $L$ should not exceed the energy of a black hole of the same size.
To establish a specific model, the length
scale $L$ has to be specified.
Applied to the dynamics of the Universe, $L$ has to be a cosmological length scale. Three different choices of $L$ have been considered in the literature:
the Hubble scale, the future event horizon and a scale proportional to the inverse square root of the Ricci scalar. Each of these choices will be discussed separately in the subsequent sections. All relations of the present section are generally valid.

Differentiating the expression (\ref{ans}) and inserting the result into the balance (\ref{cons2}) yields
\begin{equation}
\Gamma \equiv \frac{Q}{\rho_{H}} = 2\frac{\dot{L}}{L} - 3H\left(1+w\right) \ , \label{QL}
\end{equation}
where $\Gamma$ denotes the (generally time-varying) rate by which $\rho_{H}$ changes as a result of the interaction.
From the outset, there is no reason for $Q$ to vanish. Putting $Q=0$ establishes a specific relationship between $w$ and the ratio of the rates $\frac{\dot{L}}{L}$ and $H$. Any non-vanishing $Q$ will modify this relationship. We shall investigate here the general dynamics and afterwards consider the non-interacting limits as special cases for each of the choices for $L$.

With $Q$ from (\ref{QL}) the dynamics (\ref{dr2}) of the energy density ratio $r$ becomes
\begin{equation}
\dot{r} = - 3H\,\left(1 + r\right)\,\left[1 + \frac{w}{1+r} - \frac{2}{3} \frac{\dot{L}}{H L}\right]\ . \label{drL}
\end{equation}
The interaction free limit is characterized by (cf. eq.~(\ref{dr2}))
\begin{equation}
Q = 0 \quad \Rightarrow\quad \dot{r} = \frac{1}{1+r}\left(2\frac{\dot{L}}{L} - 3 H\right)
= 3 H\,r\,w \label{QL0}
\end{equation}
with a generally time-dependent $w$.
Different choices of the cutoff scale $L$ result in different
expressions for the total effective EoS parameter $w^{eff}$
in (\ref{w}) and in different relations between $w$ and $r$. It
will turn out that for a  Hubble-scale cutoff the ratio $r$ is
necessarily constant,while any, generally time varying, $w\neq 0$
requires $Q\neq 0$. For both the other choices, the future event horizon and the Ricci-scale cutoffs,  there exist
relations between $w$ and $r$. In particular, in both these cases a
constant ratio $r$ requires a  constant EoS parameter $w$.
In the following sections the three choices for $L$ are treated separately.

\section{Hubble-scale cutoff}
\label{hubble}

\subsection{General relations}

For $L=H^{-1}$ the holographic DE density is
\begin{equation}
\rho_H= 3\,c^2\,M_p^2 \,H^{2}  \ . \label{rh}
\end{equation}
Differentiating this expression and applying Eq.~(\ref{dH}) yields
\begin{equation}
\dot{\rho}_H= -3 H \left(1 + \frac{w}{1+r}\right)\rho_H \ . \label{drho1}
\end{equation}
Consequently,
\begin{equation}
\dot{\rho}_{H} + 3 H (1 + w)\rho_{H} = 3H w \frac{\rho_{m}}{1 + r}\, . \label{cons2l}
\end{equation}
This means, the source term $Q$, given by the right-hand side of the balance equation (\ref{cons2}), is fixed:
\begin{equation}
Q = - 3H w^{eff} \rho_{m}\, \quad \Leftrightarrow\quad w^{eff} = - \frac{Q}{3H\rho_{m}} . \label{weffH}
\end{equation}
This constitutes a direct relation between the EoS parameter and the so far unspecified interaction quantity $Q$. An interaction with $Q>0$ is essential to have a negative EoS parameter. Without interaction there exists only the solution with $w=0$, i.e., for the interaction-free case this cannot be a dark-energy model \cite{HDE}.

Combining (\ref{dr2}) and (\ref{weffH}) (or (\ref{w})), we find that necessarily $r = $ const. Any interacting holographic dark-energy model based on (\ref{rh}) is characterized by a constant ratio of the energy densities.
We emphasize that a constant value of $r$ does not imply that the EoS parameter is constant as well.
The ratio $\frac{Q}{3H\rho_{m}}$ remains unfixed. It can be freely chosen.
Combining the Friedmann equation (\ref{Fried1}) and the expression (\ref{rh}), we obtain the following relation between $r$ and the saturation parameter $c^{2}$:
\begin{equation}
c^{2}\left(1 + r\right) = 1 \quad \Leftrightarrow\quad r = \frac{1-c^{2}}{c^{2}}
\ . \label{cr}
\end{equation}
A specific value of $r$ fixes the saturation parameter. For $r\approx \frac{1}{3}$, i.e., 25\% DM and 75\% DE, we find $c^{2} \approx \frac{3}{4}$.
For the (generally time varying) deceleration parameter one has
\begin{equation}
q = - 1 - \frac{\dot{H}}{H^{2}} = \frac{1}{2}\left(1 - \frac{Q}{H \rho_{m}}\right)
\ . \label{q}
\end{equation}
In the interaction-free case we recover the Einstein - de Sitter value $q=\frac{1}{2}$.
The condition for accelerated expansion is $Q > H\rho_{m}$. To describe a transition from decelerated to
accelerated expansion, $Q$ has to change from $Q < H\rho_{m}$ to $Q > H\rho_{m}$.

\subsection{Variable EoS parameter}

The freedom in the choice of $\frac{Q}{H \rho_{m}}$ can be used to establish a viable cosmological model.
We assume (cf. \cite{HDE}) that the interaction rate $\Gamma$ is proportional to a power of the Hubble rate, equivalent to
\begin{equation}
\frac{\Gamma}{3 H r} = \mu \left(\frac{H}{H_{0}}\right)^{-n}
\qquad \Rightarrow\qquad \dot{\rho} + 3 H\left(1 -
\mu\left(\frac{H}{H_{0}}\right)^{-n}\right)\rho = 0\ . \label{ansH}
\end{equation}
The quantity $\mu$ is an interaction constant. Different
interaction rates are characterized by different values of $n$. A
growth of the parameter $\frac{Q}{3 H \rho_{m}}$ is obtained for $n
> 0$. In the spatially flat background the ansatz (\ref{ansH})
corresponds to a \textit{total} equation of state parameter
\begin{equation}
w^{eff} =\frac{p}{\rho} = - \mu \left(\frac{H}{H_{0}}\right)^{-n} = - \mu \left(\frac{\rho}{\rho_{0}}\right)^{-n/2}\
. \label{eoszeroo}
\end{equation}
The energy density (for $n\neq 0$) is that of a generalized Chaplygin gas \cite{Bento}:
\begin{equation}
\rho = \rho_{0}\left[\mu + \left(1 -
\mu\right)a^{-3n/2}\right]^{2/n} \quad \Rightarrow\quad H = H_{0}\left[\mu + \left(1 -
\mu\right)a^{-3n/2}\right]^{1/n}\ .
\label{rhosolve}
\end{equation}
Since from (\ref{q}) and (\ref{ansH}) it follows that
\begin{equation}
\mu = \frac{1}{3}\left(1 - 2q_{0}\right)
\ ,
\label{muq}
\end{equation}
where $q_{0}$ is the present value of the deceleration parameter, the Hubble rate can also be written as
\begin{equation}
\frac{H}{H_{0}} = \left(\frac{1}{3}\right)^{1/n}\left[1 - 2q_{0} + 2\left(1 +
q_{0}\right)a^{-3n/2}\right]^{1/n}\ .
\label{Hq}
\end{equation}
For the special case $n=2$, this expression is similar to that for the $\Lambda$CDM model.
The results of a statistical analysis are given in subsection \ref{hubbleanalysis}, encoded in figures \ref{hubble0.25} and \ref{hubble0.3}.

\subsection{Constant EoS parameter}

The simplest dynamics is obtained for a constant EoS parameter, corresponding to $n=0$ in (\ref{eoszeroo}).
With $Q = 3 \mu H\rho_{m}$ we have $\frac{w}{1 + r} = - \mu = $ const.
Under this condition the balance (\ref{cons2}) takes the form
\begin{equation}
\dot{\rho}_{H} + 3 H (1 - \mu)\rho_{H} = 0\, . \label{cons2b}
\end{equation}
To avoid an increase of $\rho_H$, i.e., a phantom behavior, we have to require $\mu < 1$.
This guarantees $\frac{w}{1+r}= - \frac{Q}{3H \rho_{m}} > -1$ for the effective EoS.
The deceleration parameter (\ref{q}) reduces to
$
q = \frac{1}{2}\left(1 - 3 \mu\right)
$.
The condition for accelerated is
$
q < 0 \  \Rightarrow\  3 \mu > 1$.
Exponential expansion is  obtained for
$
q = -1  \  \Rightarrow\  \mu = 1$.
From equation (\ref{cons2b}) the solution for the densities is
\begin{equation}
\rho_{H} \sim \rho_{m} \sim \rho \sim a^{-3\left(1 - \mu\right)}\label{solrho}
\end{equation}
with
\begin{equation}
a \sim t^{\frac{2}{3\left(1-\mu\right)}} \quad \mathrm{and}\quad H = \frac{2}{3\left(1-\mu\right)\,t} \ .\label{sola}
\end{equation}

It is well known, that any power-law solution for the scale factor corresponds to a scalar-field representation for the cosmic substratum with an effective exponential potential $V$. In the present context we have
\begin{equation}
V  = V_{0} \exp\left[\mp\lambda \left(\phi - \phi_{0}\right)\right]
\ ,
\label{Veff}
\end{equation}
with
\begin{equation}
\lambda = \sqrt{24\pi G \left(1 + r\right)\left(1 + w^{eff}\right)}
\ .
\label{leff}
\end{equation}
For $\lambda$ to be real, $w^{eff}>-1$
is required. This coincides with the condition
for avoiding a phantom-type behavior, mentioned below (\ref{cons2b}).

\section{Future event horizon cutoff}
\label{event}

\subsection{General relations}

Since accelerated expansion cannot be accounted for by holographic DE with an Hubble-scale cutoff without interaction (cf. (\ref{q})), Li \cite{li} replaced the Hubble scale by the future event horizon $R_{E}$. This approach has received considerable attention subsequently \cite{futureEH}.

With $L=R_{E}$ the holographic DE density (\ref{ans}) is
\begin{equation}
\rho_H= \frac{3\,c^2\,M_p^2}{R_{E}^{2}}  \  ,\label{rhoE}
\end{equation}
where
\begin{equation}
R_E(t)=a(t)\,\int_t^\infty\,\frac{dt'}{a(t')}=a\,\int_a^\infty\,\frac{da'}{H'\,a'^2}\
\label{re}
\end{equation}
is the future event horizon.
Differentiating (\ref{rhoE}) yields
\begin{equation}
\dot{\rho}_H= - 2 \frac{\dot{R}_{E}}{R_{E}} \rho_H\ . \label{drhoE}
\end{equation}
From (\ref{re}) we obtain
\begin{equation}
\frac{\dot{R}_E}{R_E}=H-\frac{1}{R_E}.\label{dre}
\end{equation}
Combining relations (\ref{drhoE}) and (\ref{dre}) results in
\begin{equation}
\dot{\rho}_H= - 2 \left(H - R_{E}^{-1}\right) \rho_H\ . \label{drho1E}
\end{equation}
For the left-hand side of the balance (\ref{cons2}) we obtain
\begin{equation}
\dot{\rho}_{H} + 3 H (1 + w)\rho_{H} = \left[\left(1 + 3w\right)H + 2 R_{E}^{-1}\right]\rho_{H}\, . \label{cons22}
\end{equation}
Comparing this equation with (\ref{cons2}), we may read off
\begin{equation}
Q = - H\left[1 + 3w + \frac{2}{R_{E} H}\right]\rho_{H}\, \quad \Leftrightarrow \quad
w=-\frac{1}{3}\,\left[1+\,\frac{2}{c\,\sqrt{1+r}}\,+\frac{\Gamma}{H}\right]\ .\
\label{QE}
\end{equation}
Here we have used that $R_E=c\sqrt{(1+r)}/H$, a relation that allows us to eliminate $c$ and to write
\begin{equation}
R_{E}H = R_{E0}H_{0}\sqrt{\frac{1+r}{1+r_{0}}} \quad \Leftrightarrow\quad
r = r_{0} + \left(1 + r_{0}\right)\left[\frac{R_{E}^{2}H^{2}}{R_{E0}^{2}H^{2}_{0}} - 1\right]
\ .\label{REH}
\end{equation}
Notice that $r_{0} = \frac{\Omega_{m0}}{1 - \Omega_{m0}}$, where $\Omega_{m0} = \frac{8\pi G\rho_{m0}}{3H_{0}^{2}}$.
To have an energy transfer from DE to DM one has to require
\begin{equation}
Q >0 \quad \Leftrightarrow \quad w< - \frac{1}{3} \left(1 + \frac{2}{R_{E}H}\right)\ .
\label{QE>0}
\end{equation}
The dynamics of the ratio $r$ is determined by (\ref{dr2}).
With (\ref{QE}) in (\ref{dr2}) we may also write
\begin{equation}
\dot{r} = - Hr\,\left[1 + 2 \left(R_{E}H\right)^{-1} - \frac{Q}{H\rho_{m}}\right]\ , \label{dr3}
\end{equation}
or
\begin{equation}
\dot{r} = - H\left(1 + r\right)
\left[1 + 3\frac{w}{1+r} + \frac{2}{R_{E}H}\right]\ . \label{drfh}
\end{equation}
One expects the ratio $r$ to decay during the cosmic evolution, i.e.,
\begin{equation}
\dot{r} < 0 \quad \Leftrightarrow \quad w > - \frac{1}{3}\left(1 + r\right)
\left(1 + \frac{2}{R_{E}H}\right)\ . \label{dr<}
\end{equation}
This amounts to a lower bound for the EoS parameter.
Combination of the upper bound (\ref{QE>0}) with (\ref{dr<}) provides us with following range for the
total effective EoS:
\begin{equation}
- \frac{1}{3}
\left(1 + \frac{2}{R_{E}H}\right) < \frac{w}{1+r} < - \frac{1}{3\left(1 + r\right)}
\left(1 + \frac{2}{R_{E}H}\right)
\ . \label{rangeE}
\end{equation}
Notice that $q = \frac{1}{2}(1 + 3\frac{w}{1+r})$. For a given
$w(a)$, Eq. (\ref{drfh}) determines the ratio $r(a)$. A constant $r$
implies that $w$ is constant as well. This property is substantially different from the case of the previous section where a cutoff at the Hubble scale with a constant ratio $r$ still left room for an arbitrarily time-dependent EoS. The case of a constant $r$ for the present future-event-horizon cutoff will be considered
in  subsection \ref{scalingE} below.

The dark-energy balance (\ref{cons2}) an be written as
\begin{equation}
\dot{\rho}_{H} + 3 H (1 + w_{eff}^{E})\rho_{H} = 0
\  \label{deeff}
\end{equation}
with an effective EoS
\begin{equation}
w_{eff}^{E} = w + \frac{Q}{3H\rho_{H}} = - \frac{1}{3}\left(1 + \frac{2}{R_{E}H} \right)\ ,
  \label{weff}
\end{equation}
where the superscript E denotes ``event horizon".
This quantity does not directly depend on $w$. However, the ratio $r$ that enters $R_{E}H$ is determined by $w$ via Eq.~(\ref{drfh}). Notice also, that this effective EoS for the DE component is different
from the total effective EoS of the cosmic medium which is $\frac{w}{1+r}$. In the previous Hubble-scale-cutoff case both these quantities were identical. Via (\ref{weff}), a constant $w_{eff}^{E}$ necessarily implies a constant $r$ and vice versa.

\subsection{Scaling energy-density ratio}

To solve the dynamics of the present model, we shall assume a power-law dependence for the energy-density ratio \cite{dalal}
\begin{equation}
r = r_{0}a^{-\xi} \ .
\label{rxi}
\end{equation}
Under such circumstances we have
$\dot{r} = - \xi H r$ which, inserted into (\ref{drfh}), provides us with
\begin{equation}
w = - \frac{1}{3}\left[1 + r - \xi r + \frac{2}{c}\sqrt{1+r}\right] \ .
\label{wxi}
\end{equation}
The interaction rate that corresponds to a dynamics with (\ref{rxi}) is
\begin{equation}
\Gamma = H r\left[1 - \xi  + \frac{2}{c}\frac{1}{\sqrt{1+r}}\right] \ .
\label{Gamma}
\end{equation}
The first inequality in (\ref{rangeE}) is satisfied for any $\xi > 0$, while the second inequality in
(\ref{rangeE}), equivalent to $Q>0$, requires
\begin{equation}
\xi < 1 + \frac{2}{c\sqrt{1+r}}
\ .
\label{ineq2}
\end{equation}
For any $\xi > 1$ this implies an upper limit for the ratio $r$. The ansatz (\ref{rxi}) was introduced to approach the coincidence problem \cite{scaling}. As was argued in \cite{dalal}, any value $\xi < 3$ makes the coincidence problem less severe than in the $\Lambda$CDM model for which one has $\xi = 3$.
In the following we consider the cases $\xi =1, 2, 3$ separately.

\noindent (i)
Integration of (\ref{deeff}) with $\xi = 1$ leads to a holographic energy density
\begin{equation}
\rho_{H} = \rho_{H0}a^{-2}\left[\frac{\sqrt{r_{0} + a} + \sqrt{a}}{\sqrt{r_{0} + 1} + 1}\right]^{4/c}
\ \qquad \qquad (\xi = 1)\ .
\label{rh1E}
\end{equation}
With $\rho_{m} = r\rho_{H}$ the total energy density becomes $\rho  = (1 + r)\rho_{H}$, equivalent to a Hubble rate
\begin{equation}
H = H_{0}a^{-2}\,\frac{\sqrt{r_{0} + a}}{\sqrt{r_{0} + 1}}
\left[\frac{\sqrt{r_{0} + a} + \sqrt{a}}{\sqrt{r_{0} + 1} + 1}\right]^{2/c} \qquad \qquad (\xi = 1)
\ .
\label{hubbleE1}
\end{equation}
The free parameters are $H_{0}$, $r_{0} = \frac{\Omega_{0}}{1 - \Omega_{0}}$, $\xi$ and $c$.

\noindent
(ii)
The corresponding relations for $\xi = 2$ are
\begin{equation}
\rho_{H} = \rho_{H0}a^{-2}\left[\frac{\sqrt{r_{0} + a^{2}} + a}{\sqrt{r_{0} + 1} + 1}\right]^{2/c}
\  \qquad \qquad (\xi = 2)
\label{rh2E}
\end{equation}
and
\begin{equation}
H = H_{0}a^{-2}\,\frac{\sqrt{r_{0} + a^{2}}}{\sqrt{r_{0} + 1}}
\left[\frac{\sqrt{r_{0} + a^{2}} + a}{\sqrt{r_{0} + 1} + 1}\right]^{1/c}  \qquad \qquad (\xi = 2)
\ .
\label{hubbleE2}
\end{equation}

\noindent
(iii)
For  $\xi = 3$ one has
\begin{equation}
\rho_{H} = \rho_{H0}a^{-2}\left[\frac{\sqrt{r_{0} + a^{3}} + \sqrt{a^{3}}}{\sqrt{r_{0} + 1} + 1}\right]^{4/(3c)}
\ \qquad \qquad (\xi = 3)
\label{rh3E}
\end{equation}
and
\begin{equation}
H = H_{0}a^{-2}\,\frac{\sqrt{r_{0} + a^{3}}}{\sqrt{r_{0} + 1}}
\left[\frac{\sqrt{r_{0} + a^{3}} + \sqrt{a^{3}}}{\sqrt{r_{0} + 1} + 1}\right]^{2/(3c)} \qquad \qquad (\xi = 3)
\ .
\label{hubbleE3}
\end{equation}
\noindent
With the observationally preferred values for $c^{2}$ and $r_0$ (see table \ref{tab2} in section \ref{statanalysis} below),  the condition (\ref{ineq2}) is satisfied for all the three cases, i.e., the energy transfer at the present time is always from DE to DM.
But these cases are not only quantitatively but also qualitatively different. The case $\xi = 1$ has the best-fit value $c^{2} = 1.14$. Together with the preferred value $r_0 = 0.25$ this corresponds to a present effective EoS parameter
$w_{eff}^{E}$ in (\ref{weff}) of $w_{eff}^{E}(a=1) \approx -0.89$. Even in the limit $a \gg 1$ the parameter  $w_{eff}^{E}$ will remain larger than $-1$ and both the energy density $\rho_{H}$ in (\ref{rh1E}) and the Hubble rate in (\ref{hubbleE1}) will decay in the long-time limit.
The situation is different for $\xi = 2$ with the preferred values $c^{2} = 0.73$ and $r_0 = 0.51$. Although the present value of the effective EoS parameter $w_{eff}^{E}(a=1) \approx -0.97$ is larger than $-1$, it will cross the phantom line for $a>1$. Both the energy density in eq.~(\ref{rh2E}) and the Hubble rate in (\ref{hubbleE2}) increase
with $a$.
For $\xi = 3$ with $c^{2} = 0.42$ and $r_0 = 0.95$, the EoS parameter (\ref{weff}) has a present value of
$w_{eff}^{E}(a=1) \approx - 1.07$, i.e., it is of the phantom type today and will remain so in the future.
Also here, both the energy density $\rho_{H}$ in (\ref{rh3E}) and the Hubble rate (\ref{hubbleE3}) are increasing
functions of the scale factor.

A statistical analysis provides us with figures \ref{xi1}, \ref{xi2} and \ref{xi3}
in subsection \ref{ehanalysis} below.
Notice that all models with a cutoff at the future event horizon suffer from the serious drawback that they cannot describe a transition from decelerated to accelerated expansion. A future event horizon does not exist during the period of decelerated expansion. This drawback manifests itself here also in the inequality (\ref{ineq2}) which is clearly violated if $r$ is larger than a certain threshold value. In other words, a matter dominated era with $r \gg 1$ cannot be recovered in this context.
None of the Hubble rates (\ref{hubbleE1}), (\ref{hubbleE2}) and (\ref{hubbleE3}) approaches a matter-type behavior $H \propto a^{-3/2}$ for $a\ll 1$.

\subsection{Non-interacting limit}

\begin{figure}[th]
\includegraphics[width=5.0in,angle=0,clip=true]{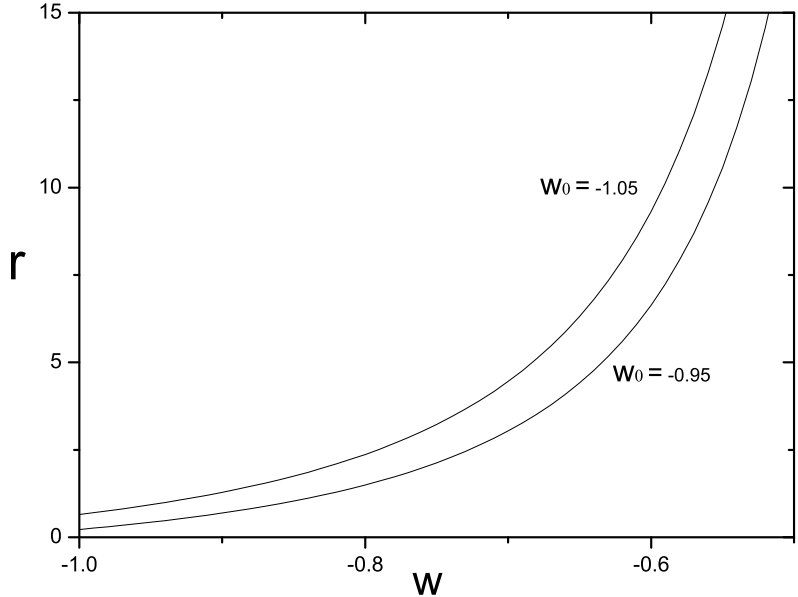}
\caption{Event-horizon cutoff, non-interacting limit. The plot shows  the ratio  $r$ versus the parameter $w$
for two  values of $w_0$  with $r_0=3/7$.
\label{fig1}}
\end{figure}

\begin{figure}[th]
\includegraphics[width=5.0in,angle=0,clip=true]{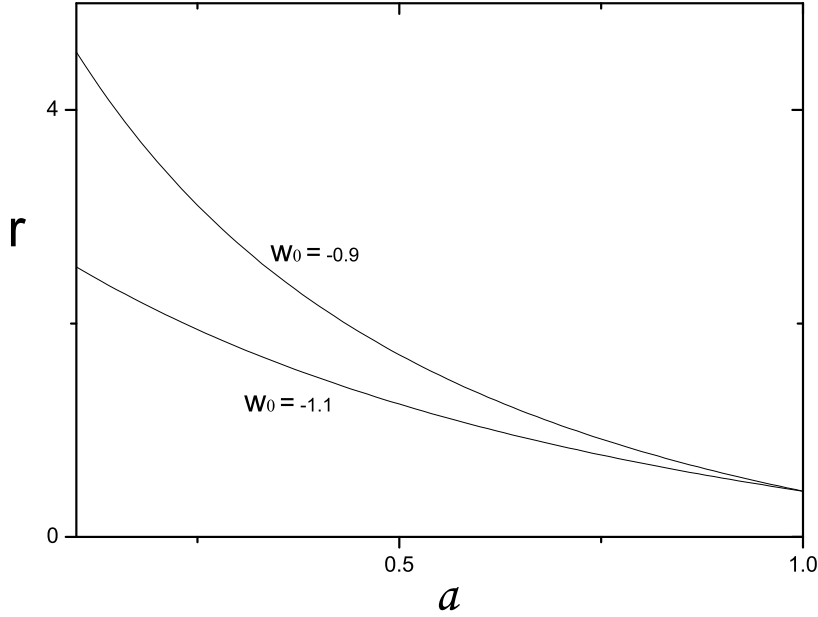}
\caption{Event-horizon cutoff, non-interacting limit. The plot shows  the ratio  $r$ versus the scale factor
$a$ for two values of $w_0$  for
$r_0=3/7$ and $w_1=0.31$. \label{fig2}}
\end{figure}

Different from the previous Hubble-scale cutoff, there exists a non-interacting limit with accelerated expansion in the present case.
In this special situation Eq.~(\ref{QE}) reduces to a relation $w=w(r)$.
With $Q=0$ one has from (\ref{QE}) that
\begin{equation}
R_{E}H = - \frac{2}{1 + 3 w}\ , \qquad R_{E0}H_{0} = c\sqrt{1+r_{0}} = - \frac{2}{1 + 3 w_{0}}
\ .
\label{Q0}
\end{equation}
For $w_{0} \approx -1$ and $r_{0} \approx \frac{1}{3}$ we have $c^{2} \approx \frac{3}{4}$ as in the Hubble-scale cutoff case.
But notice that to derive this value for the Hubble-scale cutoff, neither an assumption $Q=0$ nor
an EoS parameter $w_{0} \approx -1$ was necessary.
Together with (\ref{REH}) we obtain $r=r(w)$,
\begin{equation}
\frac{1+r}{1+r_{0}} = \left(\frac{1 + 3 w_{0}}{1 + 3 w}\right)^{2}\quad \Leftrightarrow
r = r_{0} + \left(1 + r_{0}\right)\left[\left(\frac{1 + 3 w_{0}}{1 + 3 w}\right)^{2} - 1\right]
\ .
\label{rEw}
\end{equation}
Alternatively,
\begin{equation}
w = - \frac{1}{3} + \frac{1}{3}\left(1 + 3w_{0}\right)\sqrt{\frac{1+r_{0}}{1+r}}
\ .
\label{wrE}
\end{equation}
It is obvious, that for any $w_{0}\approx -1$, the parameter $w$ remains always smaller than $-\frac{1}{3}$, demonstrating again the impossibility
of a matter-dominated period in this context. This circumstance is
visualized in a $r-w$ plane in Fig.~\ref{fig1}
for two  values of $w_0$. The corresponding dependence $r(a)$ is shown in Fig.~\ref{fig2} for the
parametrization $w = w_{0} + w_{1}(1-a)$, where $w_{0}$ and $w_{1}$ are constants \cite{CPL}.


\subsection{Scaling solution}
\label{scalingE}

According to (\ref{dr3}), a stationary solution  $r = r_{0}$ requires
\begin{equation}
\frac{Q}{H\rho_{m}} = 1 + 2 \left(R_{E}H\right)^{-1} \quad \Rightarrow\quad
\frac{w}{1 + r_{0}} = -\frac{1}{3}\left[1 + \frac{2}{R_{E0}H_{0}} \right]\ .\label{cond}
\end{equation}
Comparison with (\ref{weff}) shows, that in this case the total effective EoS parameter $\frac{w}{1 + r}$ coincides with the effective EoS parameter $w_{eff}^{E}$.
Notice, that $\dot{r} = 0$ is a separate requirement here. For the Hubble-scale cutoff a constant $r$ was a property of the model.
A further difference is the following. While for the Hubble-scale cutoff the constant ratio $r$ was compatible with a time-dependent equation of state, the EoS parameter $w$ is necessarily constant for the present configuration. It is completely determined by $r$ and the saturation parameter $c$. A similar property holds for the ratio $\frac{Q}{H\rho_{m}}$ in (\ref{cond}). This ratio is fixed as well. Recall, that for the Hubble-scale cutoff the corresponding quantity could be chosen freely.
Notice that according to the last relation in (\ref{cond},) the saturation parameter is given both
by $r_{0}$ and $w_{0}$. For the Hubble scale cutoff it was given by $r=r_{0}$, independently of the EoS parameter.

For constant  $r$ the energy densities behave as
\begin{equation}
\rho \propto \rho_{m} \propto \rho_{H} \propto a ^{-2\left[1 - \left(R_{E0}H_{0}\right)^{-1}\right]}\, . \label{solrh}
\end{equation}
Integrating Friedmann's equation, we obtain the power-law solution
\begin{equation}
a \propto t^{\frac{1}{1- 1/(R_{E0}H_{0})}}\, , \quad\rightarrow\quad H = \frac{1}{1 - 1/(R_{E0}H_{0})}\frac{1}{t}\ . \label{a}
\end{equation}
The deceleration parameter becomes
\begin{equation}
q = - 1 - \frac{\dot{H}}{H^{2}} = - \frac{1}{R_{E0}H_{0}} = - \frac{1}{c\sqrt{1+r_{0}}}\ . \label{qE}
\end{equation}
As expected, $q$ is always negative. For holographic dark energy with an infrared cutoff at the future event horizon, a stationary solution for $r$ is not compatible with a matter
dominated phase. We recall that an event horizon does not exist for decelerated expansion, i.e., it is impossible to describe a transition from decelerated to accelerated expansion within a model with future event horizon cutoff.

The equivalent scalar field potential for this case coincides with (\ref{Veff}) and
(\ref{leff}) where now $w^{eff}$ is determined by $\frac{w}{1+r_{0}}$
from (\ref{cond}).

\section{Ricci scale cutoff}
\label{riccicut}

\subsection{General relations}

The role of a distance proportional to the Ricci scale as a causal connection scale for
perturbations was noticed in \cite{brustein}. As a cutoff length in DE
models it was first used in \cite{021}. Observational constraints
were obtained, e.g., in \cite{xu,cai,zhang,xulu,suma,yi,xuwang,ivdi,rong}. The Ricci scalar
is given by $R = 6\left(2H^{2} + \dot{H}\right)$.
The corresponding cutoff-scale quantity is $L^{2} = \frac{6}{R}$, i.e.,
\begin{equation}
\rho_H= 3\,c^2\,M_p^2 \,\frac{R}{6} = \alpha\left(2H^{2} + \dot{H}\right) \ ,  \label{rhR}
\end{equation}
where $\alpha = \frac{3c^{2}}{8\pi G}$.
Upon using (\ref{dH})
we obtain the expression
\begin{equation}
\rho_H= \frac{\alpha}{2}\,H^{2}\left(1 - 3\frac{w}{1+r}\right)  \label{rh2}
\end{equation}
for the holographic dark-energy density.
It is remarkable, that the EoS parameter explicitly enters $\rho_H$. With $3H^{2} = 8\pi G \rho_{H}\left(1+r\right)$ we find
\begin{equation}
1 =  \frac{c^{2}}{2}\left(1 + r - 3 w\right) \quad \Rightarrow\quad w = \frac{1}{3}\left(1 + r\right) - \frac{2}{3c^{2}}\ .
\label{1=}
\end{equation}
Already at this stage it is obvious, that a constant value of $w$ necessarily implies a constant $r$ and vice versa.
The time dependence of $w$ is related to that of $r$ by $\dot{r} = 3\dot{w}$.
The last relation in (\ref{1=}) may be used to express $c^{2}$ in terms of the present values of $w$ and $r$:
\begin{equation}
\frac{2}{c^{2}} =  1 + r_{0} - 3 w_{0}\ ,\quad \Rightarrow \quad  r = r_{0} + 3(w - w_{0})\ .
\label{c=}
\end{equation}
Recall that for the Hubble scale cutoff it was only the ratio $r=r_{0}$ which determined the saturation parameter.
Here, $c$ is again related both to $r_{0}$ and $w_{0}$, although by a relation that is different from the corresponding relation for the cutoff at the future event horizon.

Differentiating (\ref{rhR}) yields
\begin{equation}
\dot{\rho}_H= \alpha \left(4H\dot{H} + \ddot{H}\right)\ . \label{drhoR}
\end{equation}
From (\ref{dH}) one obtains
\begin{equation}
\ddot{H} = \frac{9}{2}H^{3}\left(1 + \frac{w}{1+r}\right)^{2} - \frac{3}{2}H^{2}
\left(\frac{\dot{w}}{1+r} - \frac{w\dot{r}}{\left(1 + r\right)^{2}}\right)
\ . \label{ddH}
\end{equation}
Introducing (\ref{dH}) and (\ref{ddH}) in (\ref{drhoR}), the latter becomes
\begin{equation}
\dot{\rho}_H= \alpha \left[- 6 H^{3}\left(1 + \frac{w}{1+r}\right) + \frac{9}{2}H^{3}\left(1 + \frac{w}{1+r}\right)^{2} - \frac{3}{2}H^{2}\left(\frac{\dot{w}}{1+r} - \frac{w\dot{r}}{\left(1 + r\right)^{2}}\right)\right]\ . \label{drhoR+}
\end{equation}
From the definition (\ref{rhR}) with (\ref{dH}) we have
\begin{equation}
3H\left(1 + w\right)\rho_{H} = \alpha \left[6H^{3}\left(1 + w\right) - \frac{9}{2}H^{3}\left(1 + w\right)\left(1 + \frac{w}{1+r}\right)\right]
\ . \label{3Hr}
\end{equation}
Adding up the expressions  (\ref{drhoR+}) and (\ref{3Hr}) we find
\begin{equation}
\dot{\rho}_H + 3H\left(1 + w\right)\rho_{H} = - Q
\ , \label{lhs}
\end{equation}
where
\begin{equation}
Q  = - \frac{3H}{1+r}\left[r w - \frac{\dot{w}}{H}\right]\rho_{H}
\ . \label{Q=}
\end{equation}
This general relation for the interaction term is a property of the model, determined by the ansatz (\ref{rhR}). No further assumption has entered here.
The DE balance (\ref{cons2}) may then be written as $\dot{\rho}_H + 3H\left(1 + w_{eff}^{R}\right)\rho_{H} = 0$
with
\begin{equation}
w_{eff}^{R} = \frac{1}{1+r}\left(w + \frac{\dot{w}}{H}\right) = \frac{w + \frac{\dot{w}}{H}}{1+r_{0} + 3\left(w - w_{0}\right)}
\ . \label{wef}
\end{equation}
The superscript R stands for ``Ricci" and $r_{0} = \frac{\Omega_{m0}}{1 - \Omega_{m0}}$.
Recall that the total effective EoS parameter is given by (\ref{w}).
Eq. (\ref{dH}) for the present case takes the form
\begin{equation}
\frac{d\ln H}{d\ln a} = - \frac{3}{2}\frac{1 + r_{0} + 4\left(w - \frac{3}{4}w_{0}\right)}
{1 + r_{0} + 3\left(w - w_{0}\right)}
\ . \label{dHRpr}
\end{equation}
For a given dependence $w=w(a)$ the last equation provides us with $H(a)$.
The deceleration parameter $q = - 1 - \frac{\dot{H}}{H^{2}}$ becomes
\begin{equation}
q = \frac{1}{2}\frac{1 + r_{0} + 3\left(2w - w_{0}\right)}
{1 + r_{0} + 3\left(w - w_{0}\right)}
\ . \label{qR}
\end{equation}
Its present value is
\begin{equation}
q_{0} = \frac{1}{2}\frac{1 + r_{0} + 3w_{0}}
{1 + r_{0}}
\ . \label{qR0}
\end{equation}
The transition between  phases of decelerated and accelerated expansion is given by $q=0$ in (\ref{qR}).
Denoting the corresponding value of the EoS parameter by $w_{q}$ we have
\begin{equation}
w_{q} = \frac{1}{2}w_{0} - \frac{1}{6}\left(1 + r_{0}\right)
\ . \label{wq}
\end{equation}

\subsection{The CPL parametrization}

For a given dependence $w(a)$, relation (\ref{wq}) provides us with the value $a_{q}$ of the scale factor and, equivalently, the redshift at which the transition occurs. With the CPL \cite{CPL} parametrization  $w = w_{0} + w_{1}(1-a)$
we find
\begin{equation}
a_{q} = 1 + \frac{1}{2}\frac{w_{0}}{w_{1}} + \frac{1}{6}\frac{1 + r_{0}}{w_{1}}\ , \quad z_{q} = \frac{1}{a_{q}} -1
\ . \label{zq}
\end{equation}
In order to have a transition before the present time, i.e. $a_{q} < 1$, the condition
\[
\frac{1}{2}\frac{w_{0}}{w_{1}} + \frac{1}{6}\frac{1 + r_{0}}{w_{1}} < 0
\]
has to be fulfilled.
For $w_{1} > 0$ this reduces to
\[
w_{0} < -\frac{1}{3}\left(1 + r_{0}\right)\ ,
\]
which coincides with the condition for accelerated expansion at the present time (cf. Eq.~(\ref{qR0})).
For $w_{1} < 0$ we have the opposite inequality, so this case has to be excluded. With the exception of the Constitution dataset, all the other SNIa datasets as well as the CMB and BAO data prefer positive
values for $w_{1}$ \cite{dataw1}.

Positive values of $w_{1}$ are also expected from the behavior of the energy density ratio $r = r_{0} + 3(w - w_{0})$. For $a<1$ one expects $r>r_{0}$ because the matter fraction is supposed to evolve towards smaller values with increasing $a$. This is guaranteed for $w>w_{0}$ for $a<1$. For negative $w$ this means, that in the past $w$ was less negative than the present value $w_{0}$. With $w-w_{0} = w_{1}(1-a)$ this is realized for $w_{1}>0$.

A further point is the direction of the energy transfer. From (\ref{Q=}) it is obvious that for any constant negative $w$ the source term $Q$ is positive, implying a transfer from DE to DM. For a time varying $w$ the positivity of $Q$ requires $rw - a w^{\prime} < 0$. For the CLP parametrization this becomes $rw + a w_{1} < 0$. At the present time this reduces to $w_{1} < - r_{0}w_{0}$, providing us with an upper limit for $w_{1}$.

With the CPL parametrization, eq.~(\ref{dHRpr}) can directly be integrated. The resulting Hubble-function is
\begin{equation}
H = H_{0}\,a^{-\frac{3}{2}\frac{1 + r_{0} + w_{0} + 4w_{1}}{1 + r_{0} + 3w_{1}}}\,
\left[\frac{1 + r_{0} + 3w_{1}\left(1-a\right)}{1 + r_{0}}\right]^{-\frac{1}{2}\frac{1 + r_{0} - 3w_{0}}{1 + r_{0} + 3w_{1}}}
\ . \label{HCPL}
\end{equation}
The free parameters are $H_{0}$, $w_{0}$, $w_{1}$ and $r_{0}$. No assumption about $Q$ was made to obtain the expression (\ref{HCPL}).
The results of the statistical analysis are displayed in figure \ref{ricci} in the first part of subsection \ref{riccianalysis}.
Notice that for the observationally preferred data the mentioned limit $w_{1} < - r_{0}w_{0}$ seems to be violated.

\subsection{The interaction $Q=3H\beta\rho_{H}$}

\begin{figure}[th]
\includegraphics[width=5.0in,angle=0,clip=true]{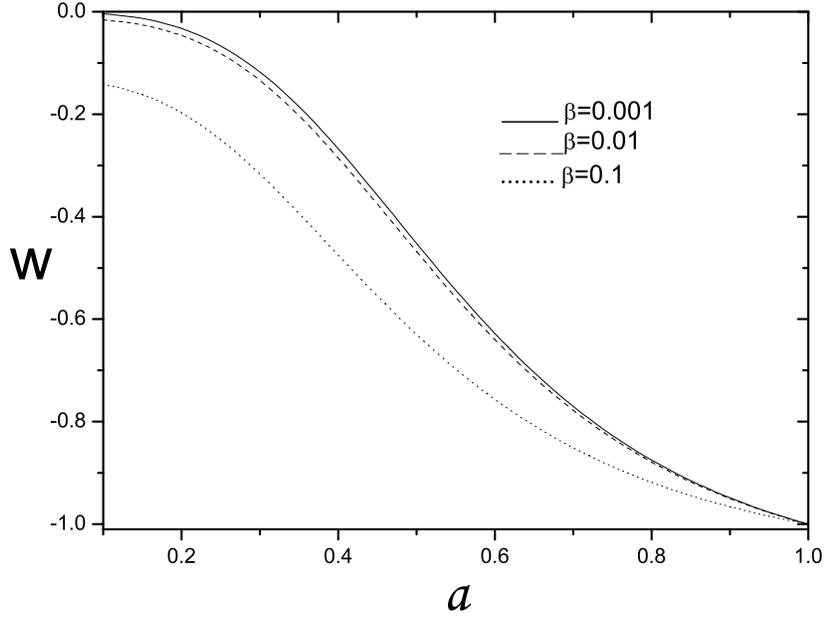}
\caption{Ricci-scale cutoff. The plot shows  the parameter $w$ versus the scale
factor $a$ for three values of the interaction parameter $\beta$  for
$w_0=-1$ and  $r_0=3/7$. \label{fig3}}
\end{figure}

\begin{figure}[th]
\includegraphics[width=5.0in,angle=0,clip=true]{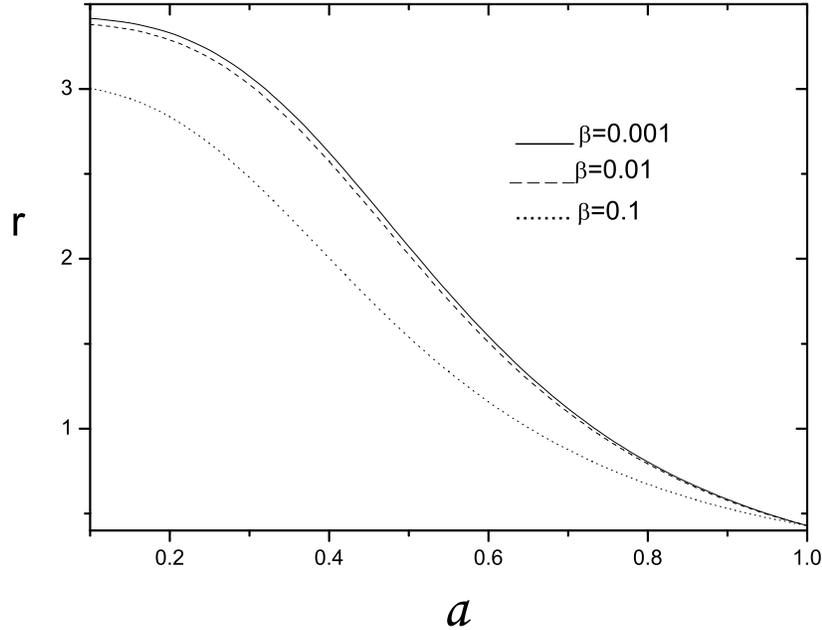}
\caption{Ricci-scale cutoff. The plot shows  the ratio  $r$ versus the scale factor
$a$ for three values of the interaction parameter $\beta$ for
$w_0=-1$ and  $r_0=3/7$. \label{fig4}}
\end{figure}

Relation (\ref{Q=}) is valid for any interaction. In this subsection we combine it with the frequently used interaction model $Q=3H\beta\rho_{H}$.
Together with the second relation of (\ref{c=}) this gives rise to a first-order differential equation for $w$ which has the solution
\begin{equation}
w = - \frac{1}{6}\frac{u-s - \left(u+s\right)\,A\,a^{s}}{1-A\,a^{s}}
\ , \label{w(a)}
\end{equation}
where
\begin{equation}
u \equiv r_{0} - 3 w_{0} + 3\beta\ ,\quad\ v \equiv r_{0} + 3 w_{0} + 3\beta\ ,\quad s\equiv \sqrt{u^{2} - 12\beta\left(1+r_{0} - 3w_{0}\right)}
\  \label{uv}
\end{equation}
and
\begin{equation}
A \equiv \frac{v-s}{v+s}
\ .  \label{A}
\end{equation}
With $w(a)$ from (\ref{w(a)}), the ratio $r(a)$ is explicitly
known as well via (\ref{c=}). In figures \ref{fig3} and
\ref{fig4} the functions $w(a)$ and $r(a)$, respectively, are shown for three
values of $\beta$. In both figures the (not included)
non-interacting limits $\beta=0$ are almost indistinguishable from
the cases $\beta=0.001$. With $w(a)$ and $r(a)$ explicitly
known, Eq.~(\ref{dH}) for the Hubble rate may be integrated. The
result is
\begin{equation}
\frac{H}{H_{0}} = a^{-\frac{3}{2}\left(1 - \frac{k}{m}\right)}\left[\frac{na^{s}-m}{n-m}\right]^{\frac{3}{2}\frac{lm-kn}{mns}}
\ ,  \label{Hint}
\end{equation}
where
\begin{equation}
m\equiv 1+r_{0}-\frac{1}{2}\left(v-s\right)\ , \quad n \equiv \left[1+r_{0}-\frac{1}{2}\left(v+s\right)\right]A
\ ,  \label{mn}
\end{equation}
\begin{equation}
k\equiv \frac{1}{6}\left(u-s\right)\ , \quad l \equiv \frac{1}{6}\left(u+s\right)A
\ .  \label{kl}
\end{equation}
The parameters here are $r_{0}$, $w_{0}$, $H_{0}$ and $\beta$.
The statistical analysis results in figure \ref{riccibis} in the second part of subsection \ref{riccianalysis}.
Observationally preferred is a small, positive value of $\beta$, i.e., a transfer from DE to DM.


\subsection{Non-interacting limit}

While, according to (\ref{Q=}), a constant EoS parameter $w$ has the non-interacting limit
$w=0$ for which $\rho_{H}$ behaves as dust, a non-interacting limit $\beta = 0$ does exist
for $\dot{w}\neq 0$. This is different from the Hubble scale cutoff case, where the non-interacting limit has $w=0$ always. In the present case
\begin{equation}
Q = 0 \quad \Rightarrow\quad r w = \frac{\dot{w}}{H}\quad \Rightarrow\quad r = \frac{d\ln w}{d\ln a}
\ . \label{rQ0}
\end{equation}
In this limit
the total effective EoS becomes
\begin{equation}
\frac{w}{1+r} = w_{0}\,\frac{r_{0}-3w_{0}}{r_{0}a^{-\left(r_{0}-3w_{0}\right)}\left[1 + r_{0}-3w_{0}\right] - 3w_{0}}
\ . \label{solwtot}
\end{equation}
It is interesting to consider the limits of $w(a)$ and $r(a)$ for small and large values of the scale factor. At high redshifts we obtain
\begin{equation}
r \rightarrow r_0 - 3w_0 \ , \qquad \frac{w}{1+r} \rightarrow 0 \qquad\qquad (a \ll 1)
\ . \label{asmall}
\end{equation}
The far-future limits are
\begin{equation}
r \rightarrow 0\ , \qquad \frac{w}{1+r} \rightarrow w_0 - \frac{1}{3}r_{0} \qquad\qquad (a \gg 1)
\ . \label{alarge}
\end{equation}
The limit (\ref{asmall}) implies that the total cosmic medium behaves as dust, thus reproducing an early matter-dominated era. Assuming tentatively $r_{0} \approx \frac{1}{3}$ and $w_{0} \approx -1$, the ratio $r$ approaches $r\approx \frac{10}{3}$ for $a \ll 1$. This value is only ten times larger than the present value $r_{0}$. Recall that for the $\Lambda$CDM model there is a difference by about nine orders of magnitude between a value of $r$ taken at the recombination epoch and its present value. In this sense, the coincidence problem is considerably alleviated for the the present model. On the other hand, in the far-future limit
$a \gg 1$ the ratio $r$ approaches zero as for the $\Lambda$CDM model. Apparently, the far-future EoS can be of the phantom type for $w_0 - \frac{1}{3}r_{0} < -1$.
The non-interacting limit $\beta=0$ of (\ref{Hint}) becomes
\begin{equation}
\frac{H}{H_{0}} = a^{-3/2}\,\sqrt{\frac{3w_{0}a^{\left(r_{0}-3w_{0}\right)}- r_{0}\left[1+r_{0}-3w_{0}\right]}
{3w_{0}- r_{0}\left[1+r_{0}-3w_{0}\right]}}
\ . \label{solHQ0}
\end{equation}

\subsection{Scaling solution}

A scaling solution is realized for (cf. Eq.~(\ref{dr2}))
\begin{equation}
\dot{r} = 0 \quad \Rightarrow\quad w_{eff}^{R} = \frac{w}{1 + r} = - \frac{Q}{3H\rho_{m}}\ . \label{wR}
\end{equation}
With (\ref{c=}) we then have $\dot{w} = 0$ as well. Under this condition $\frac{Q}{3H\rho_{m}} = \mu = $ const.
An interaction with $Q>0$ is essential to have a negative EoS parameter. The second relation in (\ref{c=}) coincides with (\ref{w}) for the Hubble scale cutoff, only that there a constant ratio $r$ does not necessarily imply that $w$ is also  constant.
For the density $\rho_H$ we then have
\begin{equation}
\rho_H= \frac{\alpha}{2}\,H^{2}\left(1 + 3\mu\right)  \ .\label{rhb}
\end{equation}
The solution for the densities  are identical with (\ref{solrho}) and (\ref{sola})
of the Hubble scale cutoff.
To avoid an increase of $\rho_H$, i.e., a phantom behavior, we have to require
$\mu < 1$. This gives rise to the inequality $c^{2}\left(1 + r_{0}\right) > \frac{1}{2}$. On the other hand,
the condition to have accelerated expansion is $\mu > \frac{1}{3}$, equivalent to $c^{2}\left(1 + r_{0}\right) < 1$, i.e., the allowed range is
\begin{equation}
\frac{1}{2} < c^{2}\left(1 + r_{0}\right) < 1
\ .
\label{limc}
\end{equation}
If we combine a specific  energy density ratio, say $r=r_{0}\approx\frac{1}{3}$ with an observationally suggested  present EoS $w \approx -1$, we obtain
$c^{2} \approx \frac{6}{13} \approx 0.46$, i.e., an exact value which is inside the range (\ref{limc}). In other words, with the observational input $r_{0}\approx\frac{1}{3}$ and $w_{0} \approx -1$ the model is completely specified. In particular, the observational data for $r_{0}$ and $w_{0}$ determine
the degree of saturation of the holographic condition (\ref{DEineq}). Notice that the value found here is  smaller than the corresponding value for the Hubble-scale cutoff.
Also for the present case the scalar field representation coincides with that for the Hubble-scale cutoff. In particular, we obtain the potential (\ref{Veff}) with (\ref{leff}).
The dependence (\ref{Veff}) with (\ref{leff}) is characteristic for all the stationary solutions.

\section{Statistical analysis and observational constraints}
\label{statanalysis}

\subsection{General description}

Each of the holographic models described in the previous sections provides us with a specific prediction for the behavior of the Hubble parameter as a
function of the scale factor.
The relevant expressions are (\ref{Hq}) for the Hubble-scale cutoff, (\ref{hubbleE1}), (\ref{hubbleE2}) and (\ref{hubbleE3}) for different cases of the event-horizon cutoff, (\ref{HCPL}) for the Ricci-scale cutoff using
the CPL parametrization and (\ref{Hint}) for the Rici-scale cutoff with an interaction proportional to the DE density.
We summarize them again as function of the
redshift $z = \frac{1}{a} -1$.
\begin{itemize}
\item Hubble radius.
\begin{equation}
H(z) = H_0\biggr(\frac{1}{3}\biggl)^n\biggr[(1 - 2q_0) + 2(1 + q_0)(1 + z)^\frac{3n}{2}\biggl]^\frac{1}{n}\ .
\label{HH}
\end{equation}
The free parameters are $H_{0}$, $q_{0}$ and $n$. In a first step, the Hubble parameter $H_0$ is determined by minimizing the three-dimensional $\chi^{2}$ function.
The remaining parameters then are $q_{0}$ and $n$, for which we perform a statistical analysis in subsection \ref{hubbleanalysis}.
\item Future event horizon with $\xi = 1$.
\begin{equation}
H(z) = H_0(1 + z)^{3/2 - 1/c}\sqrt{\frac{1 + r_0(1 + z)}{r_0 + 1}}\biggr[\frac{\sqrt{r_0(1 + z) + 1} + 1}{\sqrt{r_0 + 1} + 1}\biggl]^{2/c}\ .
\label{Hev1}
\end{equation}
Since $\xi$ is fixed, we have $H_{0}$, $r_{0}$ and $c$ as free parameters. $H_{0}$ is obtained as in the previous case. The free-parameter space then consists of $r_{0}$ and $c$.
The corresponding analysis is described in subsection \ref{ehanalysis}.
For the physical interpretation see the paragraph following Eq.~(\ref{hubbleE3}).
\item Future event horizon with $\xi = 2$.
\begin{equation}
H(z) = H_0(1 + z)^{1 - 1/c}\sqrt{\frac{1 + r_0(1 + z)^2}{r_0 + 1}}\biggr[\frac{\sqrt{r_0(1 + z)^2 + 1} + 1}{\sqrt{r_0 + 1} + 1}\biggl]^{1/c}\ .
\label{Hev2}
\end{equation}
As in the previous case, the free parameters are  $H_{0}$, $r_{0}$ and $c$.
See again subsection \ref{ehanalysis} and the paragraph following Eq.~(\ref{hubbleE3}).
\item Future event horizon $\xi = 3$.
\begin{equation}
H(z) = H_0(1 + z)^{1/2 - 1/c}\sqrt{\frac{1 + r_0(1 + z)^3}{r_0 + 1}}\biggr[\frac{\sqrt{r_0(1 + z)^3 + 1} + 1}{\sqrt{r_0 + 1} + 1}\biggl]^{2/(3c)}\ .
\label{Hev3}
\end{equation}
See subsection \ref{ehanalysis} and below Eq.~(\ref{hubbleE3}) also here.
\item Ricci scale with CPL parametrization.
\begin{equation}
H(z) = H_0(1 + z)^{\frac{3}{2}\frac{1 + r_0 + \omega_0 + 4\omega_1}{1 + r_0  + 3\omega_1}}\biggr[\frac{1 + r_0 + 3\omega_1\frac{z}{1 + z}}{1 + r_0}\biggl]^{-\frac{1}{2}\frac{1 + r_0 - 3\omega_0}{1 + r_0  + 3\omega_1}}\ .
\label{HRCPL}
\end{equation}
The free parameters of this model are $H_{0}$, $r_{0}$, $w_{0}$ and $w_{1}$. In this case, the minimum value of the four-dimensional $\chi^{2}$-function is used to determine both $H_{0}$ and $r_{0}$. Then, a two-dimensional analysis is performed for $w_{0}$ and $w_{1}$, as
described in subsubsection \ref{CPL}.
\item Ricci scale with interaction $Q = 3H\beta\rho_H$.
\begin{equation}
H(z) = H_0(1 + z)^{\frac{3}{2}\biggr(1 - \frac{k}{m}\biggl)}\biggr\{\frac{n(1 + z)^{-s} - m}{n - m}\biggl\}^{\frac{3}{2}\frac{lm - kn}{mns}}\ .
\label{HRint}
\end{equation}
Here, one has $H_{0}$, $r_{0}$, $w$ and $\beta$ as free parameters. We fix $w = - 1$ and
determine $H_{0}$ along the lines already described for the previous cases. The statistical analysis for $r_{0}$ and $\beta$ is the subject of
subsubsection \ref{ricciint}.
\end{itemize}

In order to probe the above models against observations, we consider four background tests which are directly related to the behavior of the function $H(z)$: the
supernova type Ia \cite{01}, the age of the very old galaxies leading to a direct measure of the $H(z)$ function \cite{verde},
the CMB shift parameter $R$ \cite{bond} and the baryonic acoustic oscillations $BAO$ \cite{zeldovich}.
We shall present the results for a combined analysis of these four tests.
\par
The supernova-type-Ia test is based on the luminosity distance function \cite{durrer}
\begin{equation}
D_L = (1 + z)\frac{c}{H_0}\int_0^z\frac{dz}{\sqrt{H(z)}},
\end{equation}
where $c$ is the velocity of light. The observational relevant quantity is the moduli distance, given by
\begin{equation}
\mu = m - M = 5\ln\biggr(\frac{D_L}{Mpc}\biggl) + 25,
\end{equation}
where $m$ is the apparent magnitude and $M$ is the absolute magnitude of a given supernova.
In what follows we shall use the data set of the Union2 sample \cite{amanullah}.
\par
The CMB shift parameter
\begin{equation}
R = \sqrt{\Omega_{m0}}\int_0^{z_d}\frac{dz}{H(z)}
\end{equation}
depends on the matter density parameter $\Omega_{m0}$ and the redshift at the decoupling between matter and radiation, $z_d = 1090$.
Observationally, $R = 1.725\pm0.018$ \cite{komatsu}. Baryonic acoustic oscillation ($BAO$) are characterized by the parameter
\begin{equation}
{\cal A} = \frac{\sqrt{\Omega_{m0}}}{[H(z_b)]^{1/3}}\biggr[\frac{1}{z_b}\int_0^{z_b}\frac{dz}{H(z)}\biggl]^{2/3},
\end{equation}
where $x_b = 0.35$. Observationally, ${\cal A} = 0.469\pm0.017$ \cite{eisenstein}. Another test we will use is the age of the very old galaxies that have evolved
passively. Our analysis is based on the 13 data for such objects listed in reference \cite{zhang}.
\par
For each of these observational tests we evaluate the fitting function $\chi^2$, given by
\begin{equation}
\chi^2 = \sum_{i=1}^n\frac{(\epsilon^{th}_i - \epsilon^{ob}_i)^2}{\sigma^2_i},
\end{equation}
where $\epsilon^{th}_i$ stands for a  theoretical estimation of the $ith$ data of a given quantity (moduli distance, parameters $R$ and $\cal{A}$, $H(z)$),
and $\epsilon^{ob}_i$ stands for the corresponding observational data, $\sigma_i$ being the error bar. From this statistical parameters we can construct the
probability distribution function (PDF),
\begin{equation}
P(x^f) = Ae^{-\chi^2(x^f)/2},
\end{equation}
where $A$ is a normalization factor and $x^f$ denotes the set of free parameters of the model.
The joint analysis is performed by summing up the different $\chi^2$ contributions, obtained for each of the tests separately.
A one-dimensional PDF for a given parameter can
be obtained by integrating over the remaining ones.
All one-dimensional estimations will be made at $2\sigma$ ($95\%$ confidence level), but
we will also display the $1\sigma$, $2\sigma$ and $3\sigma$ contours for the two-dimensional PDFs.
\par
In order to gauge our results, we perform at first the analysis for the $\Lambda CDM$ model. In the flat case, there are two free parameters,
$H_0$ and $\Omega_{m0}$ (or, alternatively, $\Omega_{\Lambda0}$). The best-fit value for $H_0$, based on the supernovae data, is $H_{0} = 70.05\ \mathrm{km/sec\ Mpc}$.
For our purposes, the most important quantity is $\Omega_{m0}$ for which we find
(at $2\sigma$) $\Omega_{m0} = 0.27^{+0.03}_{-0.02}$. In table \ref{tab1} we display the values for $\chi^2_{min}$ for each model and for each observational
test. The values for the parameters $R$ and $\cal{A}$ are only shown for illustrative purposes. Here, the fitting is remarkable good due to the fact that there is just
one observational point. The values in the last column of table \ref{tab1} seem to indicate that all the models are competitive. But one should keep in mind that the $\Lambda CDM$ model has
just two free parameters, whereas the number of free parameters is three or four for the other models.
In our analysis, however, we have fixed one or two of them, respectively.
The last two columns of table \ref{comp} show, how the statistical analysis penalizes the existence of additional parameters.
 In the following subsections we list the results for the different models in more detail.
\newline
\vspace{0.3cm}
\newline
\begin{table}[!t]
\begin{center}
\begin{tabular}{|c|c|c|c|c|c|}\hline
$\chi^2_{min}$&$SN$&$H(z)$&$CMB$&$BA0$&Total\\ \hline
$\Lambda CDM$&$542.71$&$8.07$&$0.0$&$1.02\times10^{-20}$&$550.78$\\ \hline
Hubble($\Omega_{m0}=0.25$)&$542.59$&$8.02$&$3.27\times10^{-15}$&$1.02\times10^{-20}$&$550.61$\\ \hline
Hubble($\Omega_{m0}=0.3$)&$542.59$&$8.02$&$1.27\times10^{-18}$&$1.61\times10^{-15}$&$550.61$ \\ \hline
Event horizon($\xi = 1$)&$544.49$&$8.53$&$1.08\times10^{-15}$&$5.08\times10^{-17}$&$553.03$\\ \hline
Event horizon($\xi = 2$)&$542.79$&$8.09$&$5.29\times10^{-17}$&$4.82\times10^{-19}$&$550.88$\\ \hline
Event horizon($\xi = 3$)&$542.95$&$8.19$&$5.26\times10^{-17}$&$3.09\times10^{-20}$&$551.15$\\ \hline
Ricci (CPL param.)&$543.44$&$8.01$&$1.37\times10^{-14}$&$2.02\times10^{-18}$&$551.45$\\ \hline
Ricci($Q=3H\beta\rho_H$)&$542.66$&$8.02$&$2.53\times10^{-15}$&$1.70\times10^{-20}$&$550.67$ \\ \hline
\end{tabular}
\end{center}
\caption{$\chi^2_{min}$ values for each of the models and different tests.}  \label{tab1}
\end{table}

\subsection{Hubble-scale cutoff}
\label{hubbleanalysis}

\begin{center}
\begin{figure}[!t]
\begin{minipage}[t]{0.3\linewidth}
\includegraphics[width=\linewidth]{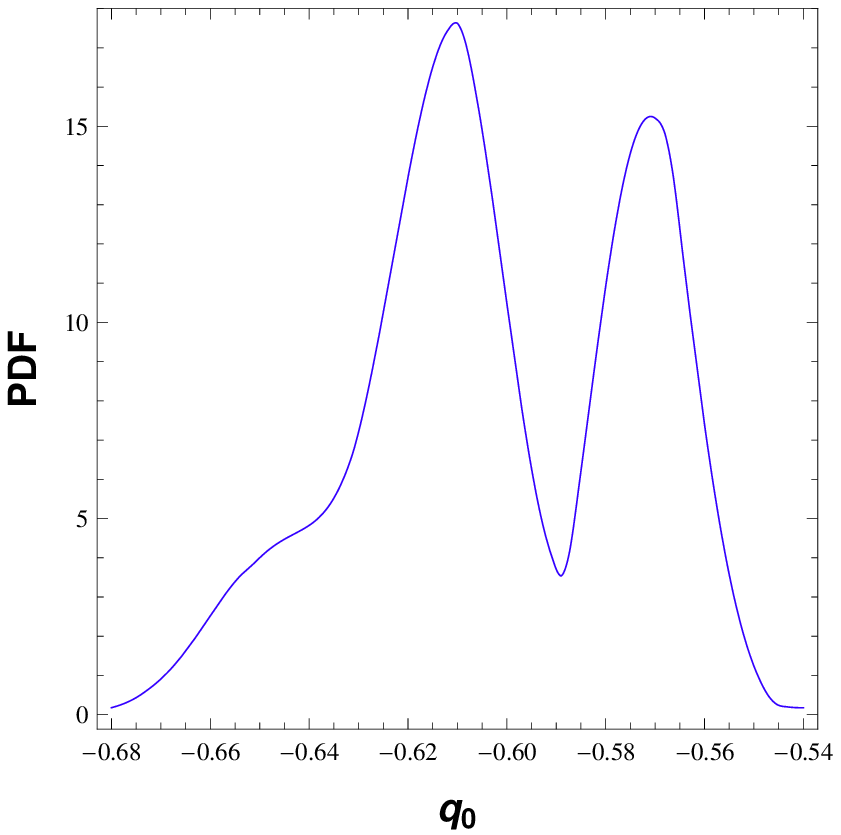}
\end{minipage} \hfill
\begin{minipage}[t]{0.3\linewidth}
\includegraphics[width=\linewidth]{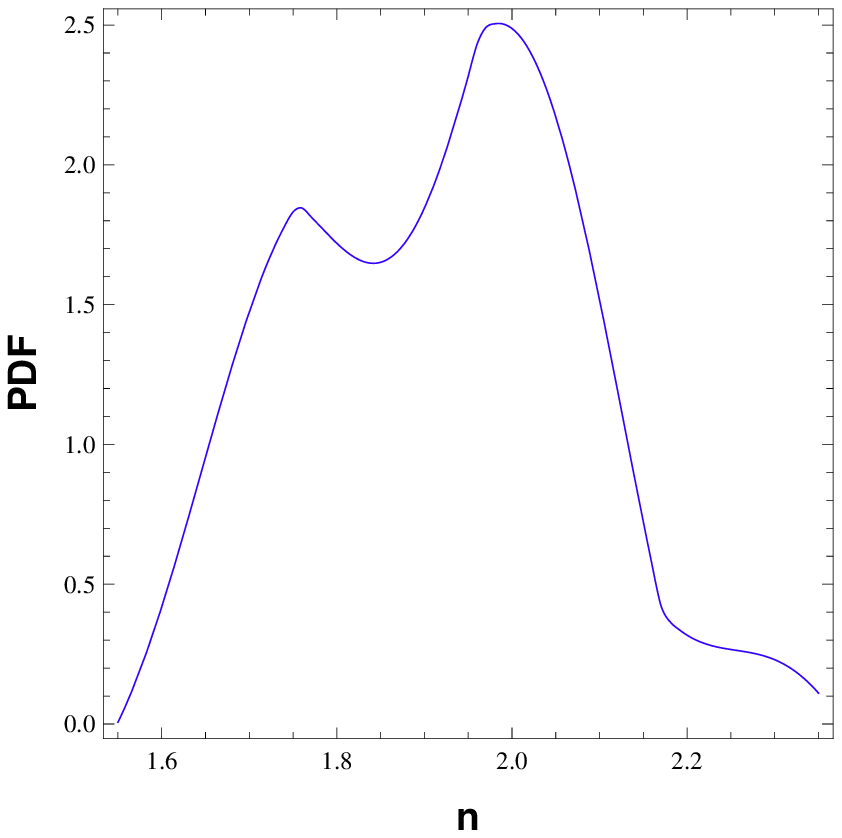}
\end{minipage} \hfill
\begin{minipage}[t]{0.3\linewidth}
\includegraphics[width=\linewidth]{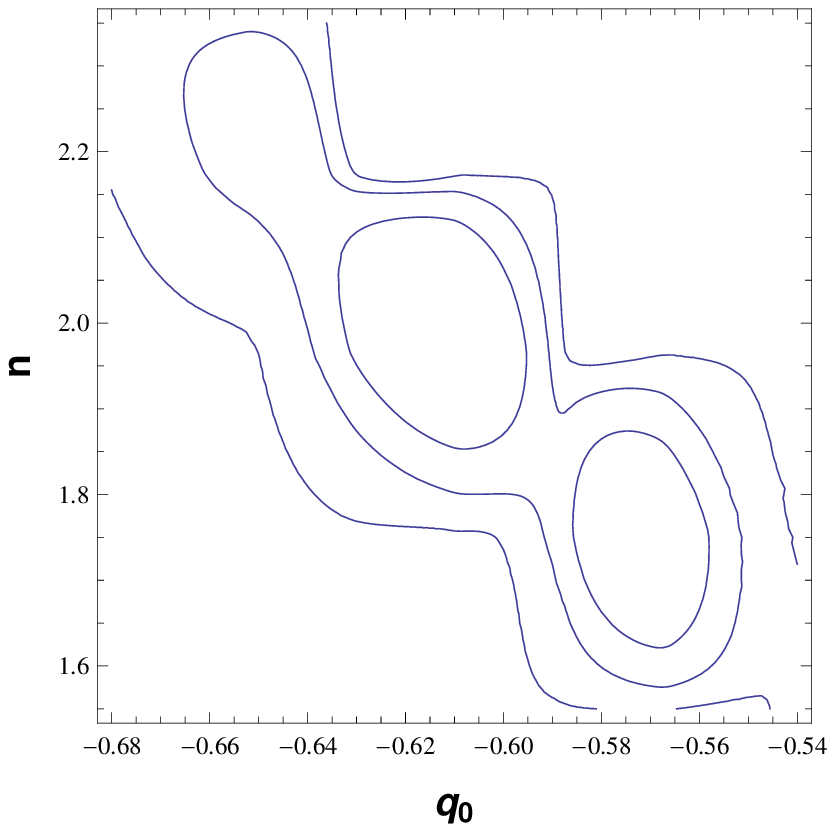}
\end{minipage} \hfill
\caption{{\protect\footnotesize Hubble scale cutoff with $\Omega_{m0} = 0.25$. Left panel: one-dimensional PDF for $q_0$. Center panel: one-dimensional PDF for $n$. Right panel: contour plots for the $1\sigma$, $2\sigma$ and
$3\sigma$ confidence levels.
}}
\label{hubble0.25}
\end{figure}
\end{center}

\begin{center}
\begin{figure}[!t]
\begin{minipage}[t]{0.3\linewidth}
\includegraphics[width=\linewidth]{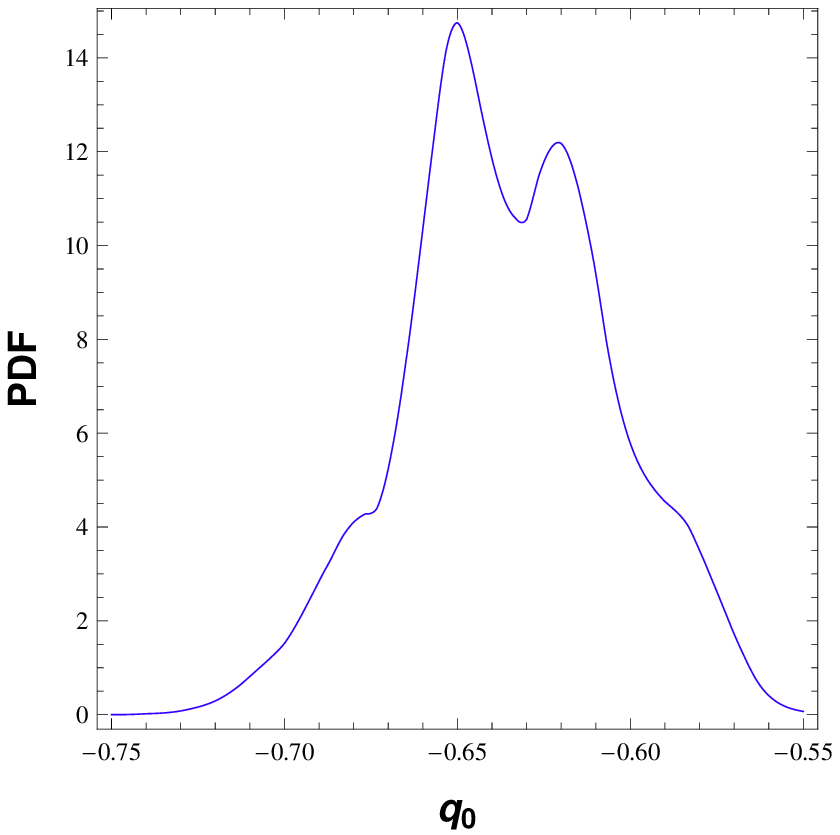}
\end{minipage} \hfill
\begin{minipage}[t]{0.3\linewidth}
\includegraphics[width=\linewidth]{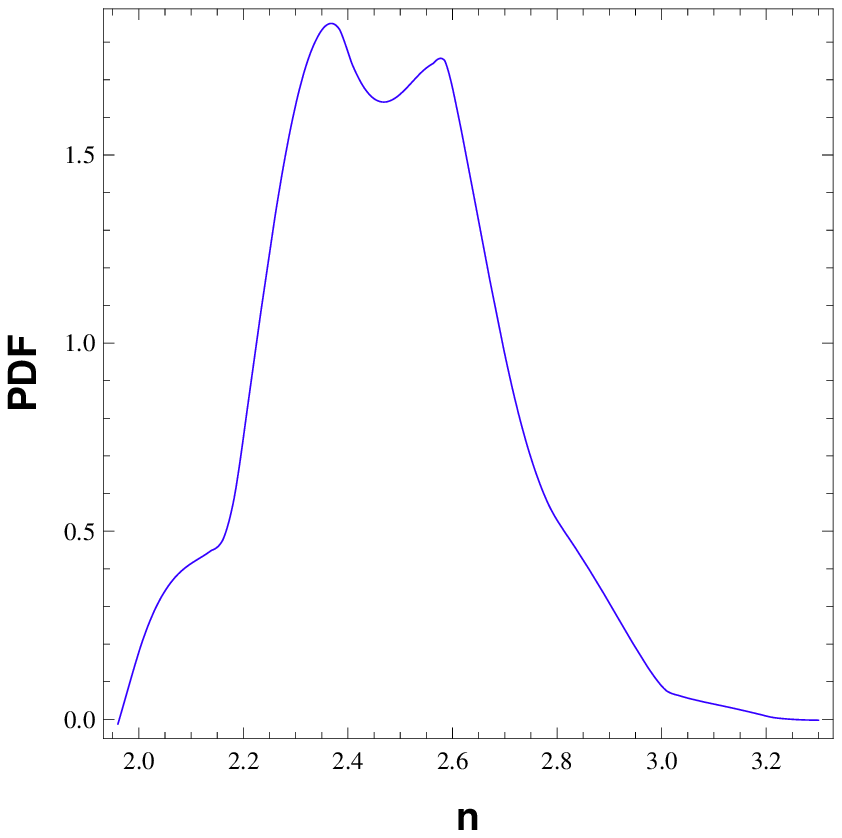}
\end{minipage} \hfill
\begin{minipage}[t]{0.3\linewidth}
\includegraphics[width=\linewidth]{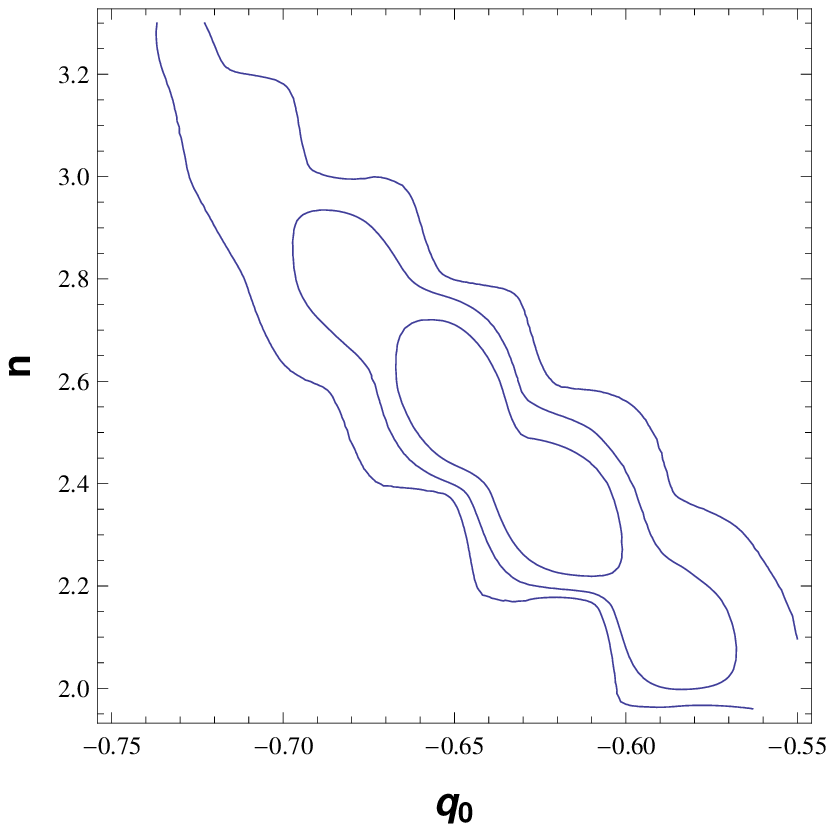}
\end{minipage} \hfill
\caption{{\protect\footnotesize Hubble scale cutoff with $\Omega_{m0} = 0.3$. Left panel: one-dimensional PDF for $q_0$. Center panel: one-dimensional PDF for $n$. Right panel: Contour plots for the $1\sigma$, $2\sigma$ and
$3\sigma$ confidence levels.
}}
\label{hubble0.3}
\end{figure}
\end{center}

In this case, there are three free parameters: $H_0$, $q_0$ and $n$ (cf.~Eq.~(\ref{HH})). The mass density has no direct connection with these parameters. This is important to stress since the CMB shift $R$ and the $BAO$ parameter ${\cal A}$ explicitly depend on the present mass density $\Omega_{m0}$.
The evaluation has been made for the cases $\Omega_{m0} = 0.25$ and $\Omega_{m0} = 0.3$. We minimize the three-dimensional $\chi^2$ function for a given set of observational data and use the value for
$H_0$ obtained in this way in order to reduce the space of parameters to two.
This procedure is appropriate, since the PDF for $H_0$ is sharply peaked around its maximum value.
It will also be applied to the other cases to be studied in the following subsections.
Generically, $H_{0} = 70\  \mathrm{km/sec\ Mpc}$ for the  supernovae data and
$H_{0} = 68\  \mathrm{km/sec\ Mpc}$ for the $H(z)$ data. 
For $\Omega_{m0} = 0.25$ the results for the remaining parameters are $q_0 = - 0.61^{+0.06}_{-0.04}$ and $n = 1.98^{+0.20}_{-0.49}$, while they are
$q_0 = -0.68^{+0.09}_{-0.02}$ and $n = 2.47^{+0.42}_{-0.42}$ for $\Omega_{m0} = 0.3$. The first case is very close to the $\Lambda CDM$ model.
The one- and two-dimensional PDFs for $q_0$ and $n$ are displayed
in figure \ref{hubble0.25} for $\Omega_{m0} = 0.25$ and in figure \ref{hubble0.3} for $\Omega_{m0} = 0.3$. Note  the existence of remarkable oscillations in these PDFs.

\subsection{Future event horizon cutoff}
\label{ehanalysis}

\begin{center}
\begin{figure}[!t]
\begin{minipage}[t]{0.3\linewidth}
\includegraphics[width=\linewidth]{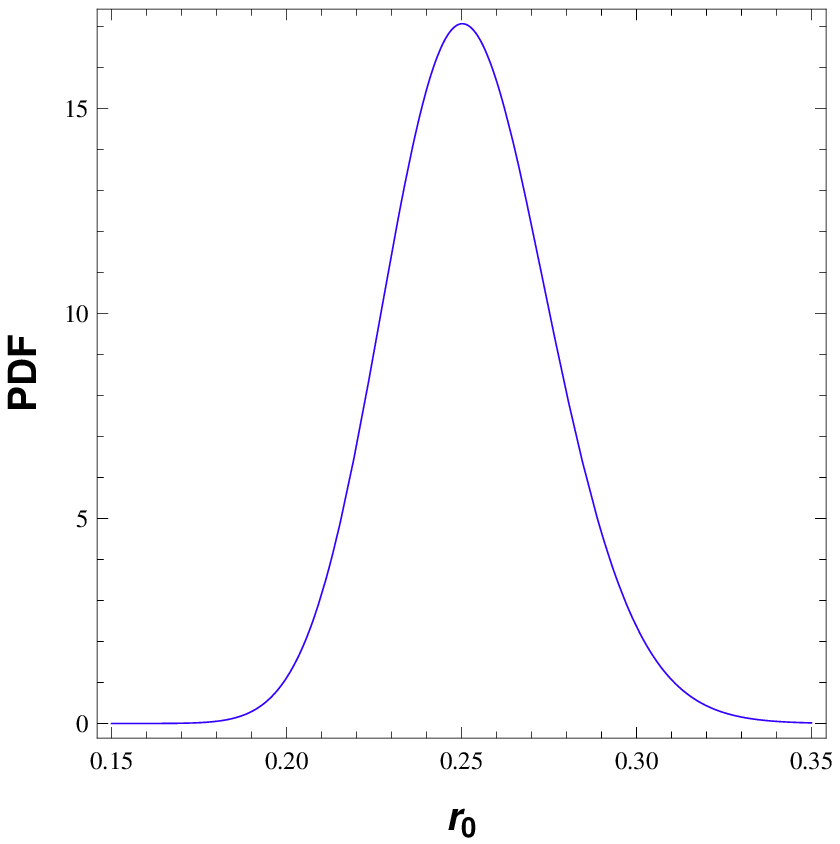}
\end{minipage} \hfill
\begin{minipage}[t]{0.3\linewidth}
\includegraphics[width=\linewidth]{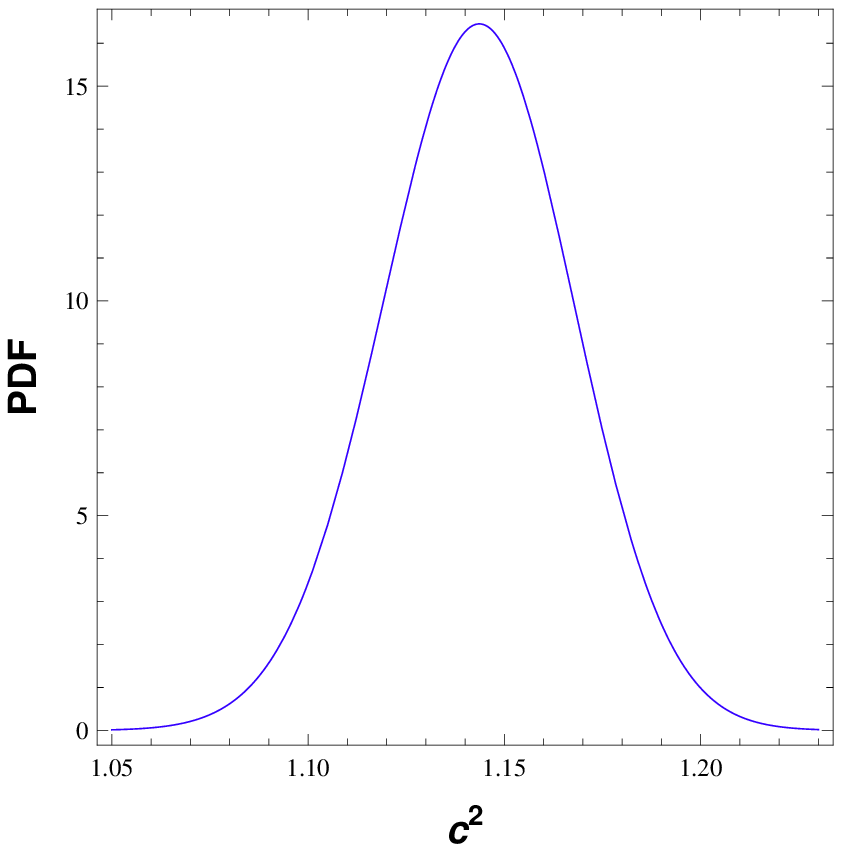}
\end{minipage} \hfill
\begin{minipage}[t]{0.3\linewidth}
\includegraphics[width=\linewidth]{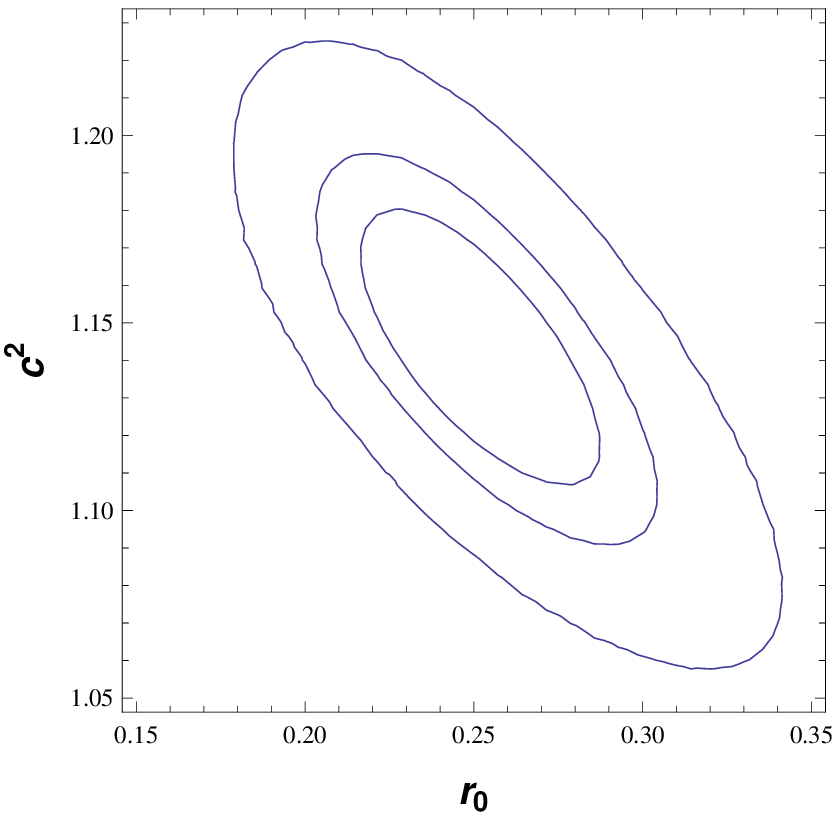}
\end{minipage} \hfill
\caption{{\protect\footnotesize Future event horizon cutoff with $\xi = 1$. Left panel: one-dimensional PDF for $r_0$. Center panel: one-dimensional PDF for $c^2$. Right panel: Contour plots for the $1\sigma$, $2\sigma$ and
$3\sigma$ confidence levels.
}}
\label{xi1}
\end{figure}
\end{center}

\begin{center}
\begin{figure}[!t]
\begin{minipage}[t]{0.3\linewidth}
\includegraphics[width=\linewidth]{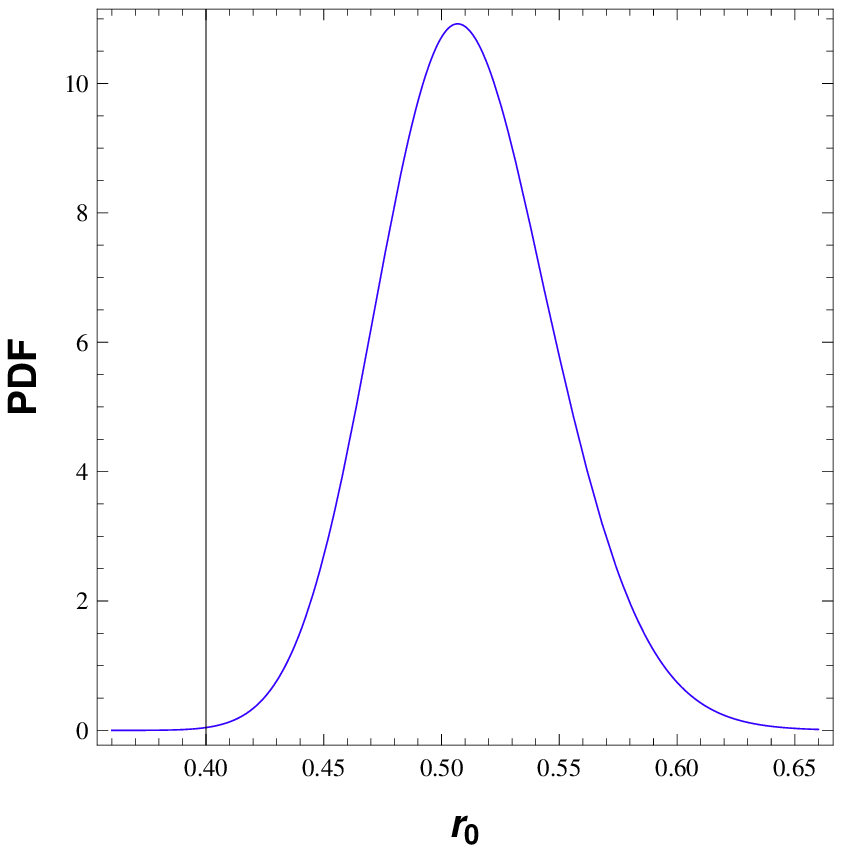}
\end{minipage} \hfill
\begin{minipage}[t]{0.3\linewidth}
\includegraphics[width=\linewidth]{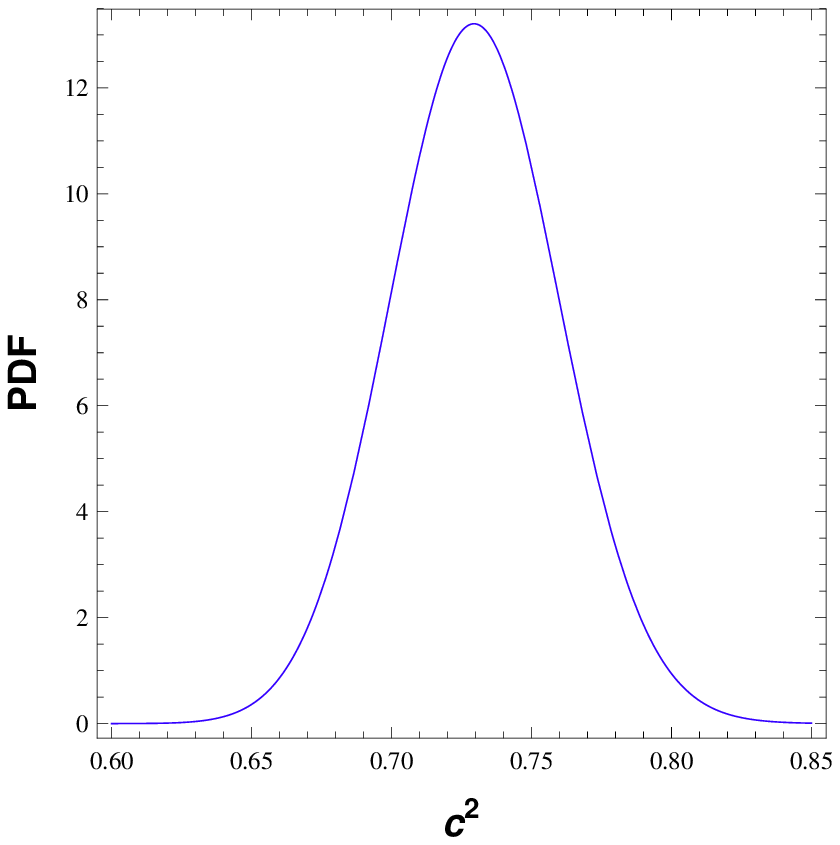}
\end{minipage} \hfill
\begin{minipage}[t]{0.3\linewidth}
\includegraphics[width=\linewidth]{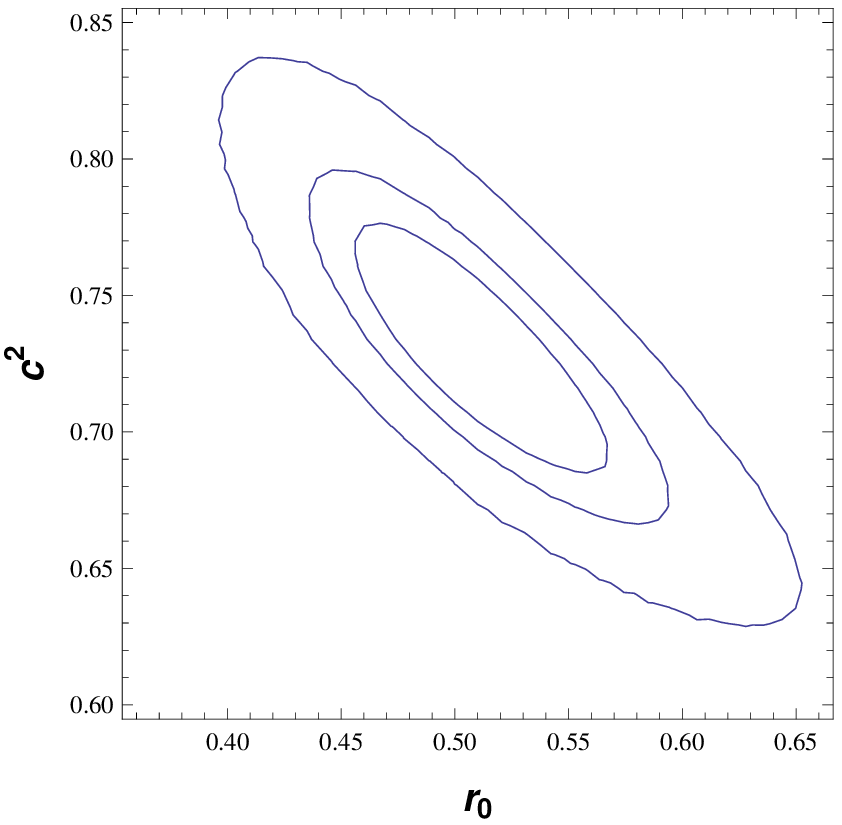}
\end{minipage} \hfill
\caption{{\protect\footnotesize Future event horizon cutoff with $\xi = 2$. Left panel: one-dimensional PDF for $r_0$. Center panel: one-dimensional PDF for $c^2$. Right panel: Contour plots for the $1\sigma$, $2\sigma$ and
$3\sigma$ confidence levels.
}}
\label{xi2}
\end{figure}
\end{center}

\begin{center}
\begin{figure}[!t]
\begin{minipage}[t]{0.3\linewidth}
\includegraphics[width=\linewidth]{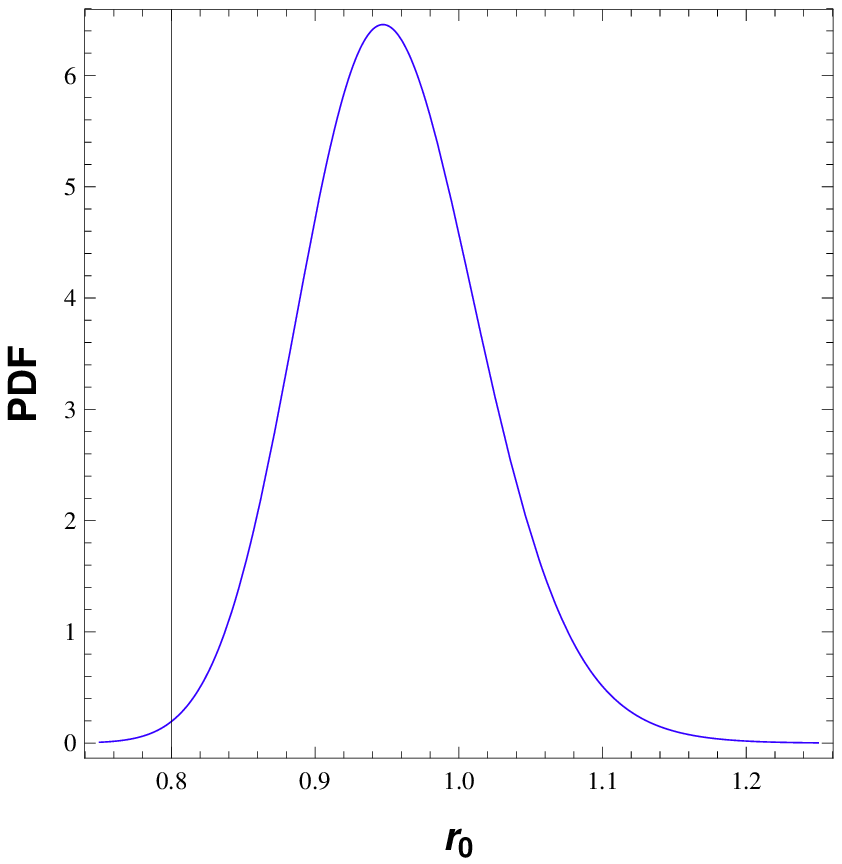}
\end{minipage} \hfill
\begin{minipage}[t]{0.3\linewidth}
\includegraphics[width=\linewidth]{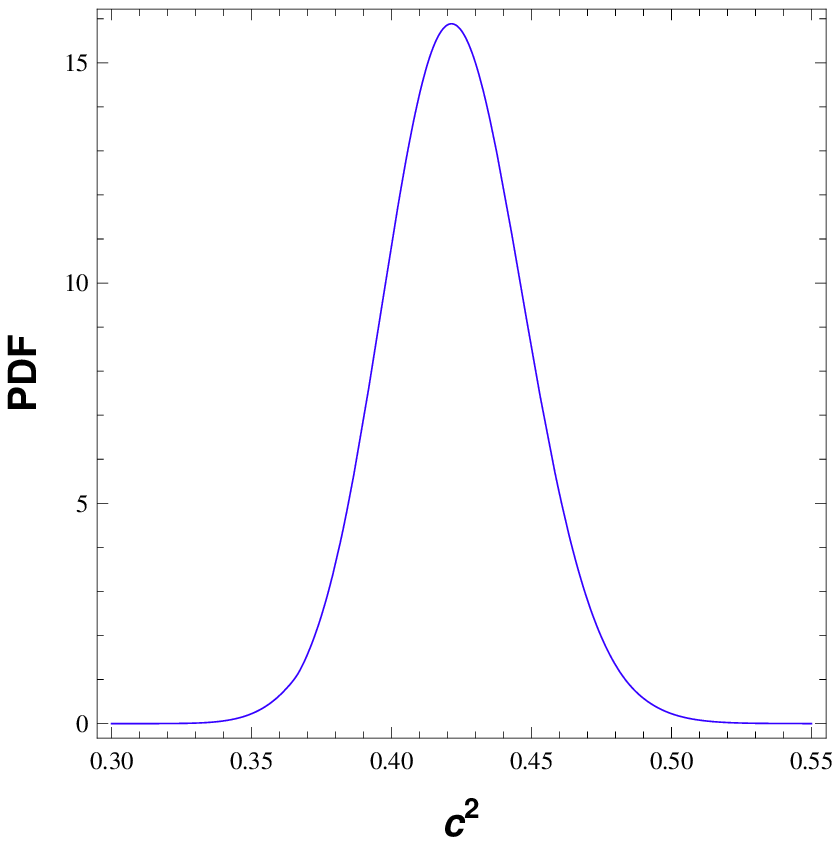}
\end{minipage} \hfill
\begin{minipage}[t]{0.3\linewidth}
\includegraphics[width=\linewidth]{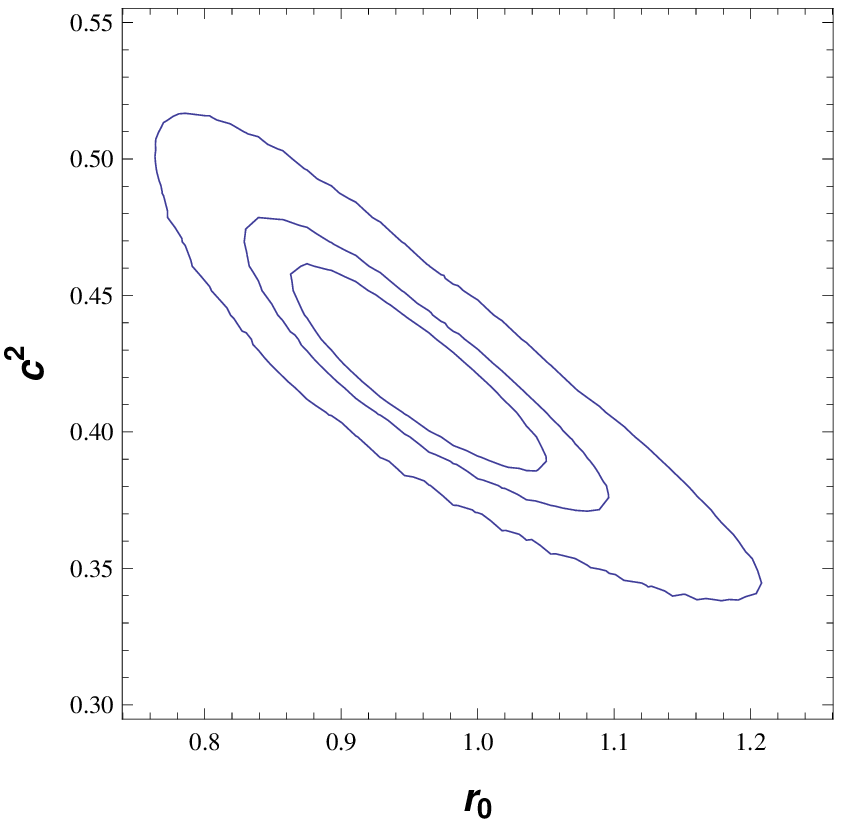}
\end{minipage} \hfill
\caption{{\protect\footnotesize Future event horizon cutoff with $\xi = 3$. Left panel: one-dimensional PDF for $r_0$. Center panel: one-dimensional PDF for $c^2$. Right panel: Contour plots for the $1\sigma$, $2\sigma$ and
$3\sigma$ confidence levels.
}}
\label{xi3}
\end{figure}
\end{center}

The free parameters are the present values of the energy density ratio $r_0$ and the Hubble rate $H_0$, as well as the saturation parameter $c$ and the power $\xi$. We use the same procedure as in the previous case in order to determine $H_0$.
Since we fix $\xi$ to be either $\xi = 1$ or  $\xi = 2$ or $\xi = 3$, we are left with two
free parameters, $r_0$ and $c$ (cf. Eqs.~(\ref{Hev1}), (\ref{Hev2}) and (\ref{Hev3})). In table \ref{tab2}, the
parameter estimations for $r_0$ and (the more convenient) $c^2$ are shown for different values of $\xi$. There is a clear tendency: as $\xi$ grows,
the density ratio increases and the value for $c^2$ decreases. In figures \ref{xi1}, \ref{xi2} and \ref{xi3} the one and two dimensional PDFs are
exhibited. The physics of each of the models has been discussed in the part following Eq.~(\ref{hubbleE3}).
Although these models are unable to reproduce a matter-dominated phase at high redshifts, the results of the statistical analysis do not differ substantially from those of the models with a correct $H \propto (1+z)^{3/2}$ behavior.
This seems to be surprising, especially as far as the CMB and BAO data are concerned. But as already mentioned,
for these tests there is just one data point and the corresponding $\chi^{2}$ values are negligible anyway.
\newline
\vspace{0.2cm}
\newline
\begin{table}[!t]
\begin{center}
\begin{tabular}{|c|c|c|c|}\hline
&$\xi = 1$&$\xi = 2$&$\xi = 3$\\ \hline
$r_0$&$0.25^{+.005}_{-0.04}$&$0.51_{-0.07}^{+0.08}$&$0.95^{+0.13}_{-0.11}$\\ \hline
$c^2$&$1.14^{+0.05}_{-0.05}$&$0.73^{+0.06}_{-0.06}$&$0.42^{+0.05}_{-0.05}$\\ \hline
\end{tabular}
\end{center}
\caption{Parameter estimations for $r_0$ and $c^2$ for different values of $\xi$.} \label{tab2}
\end{table}
\vspace{1.5cm}
\ \\

\subsection{Ricci-scale cutoff}
\label{riccianalysis}
\subsubsection{CPL parametrization}
\label{CPL}

\begin{center}
\begin{figure}[!t]
\begin{minipage}[t]{0.3\linewidth}
\includegraphics[width=\linewidth]{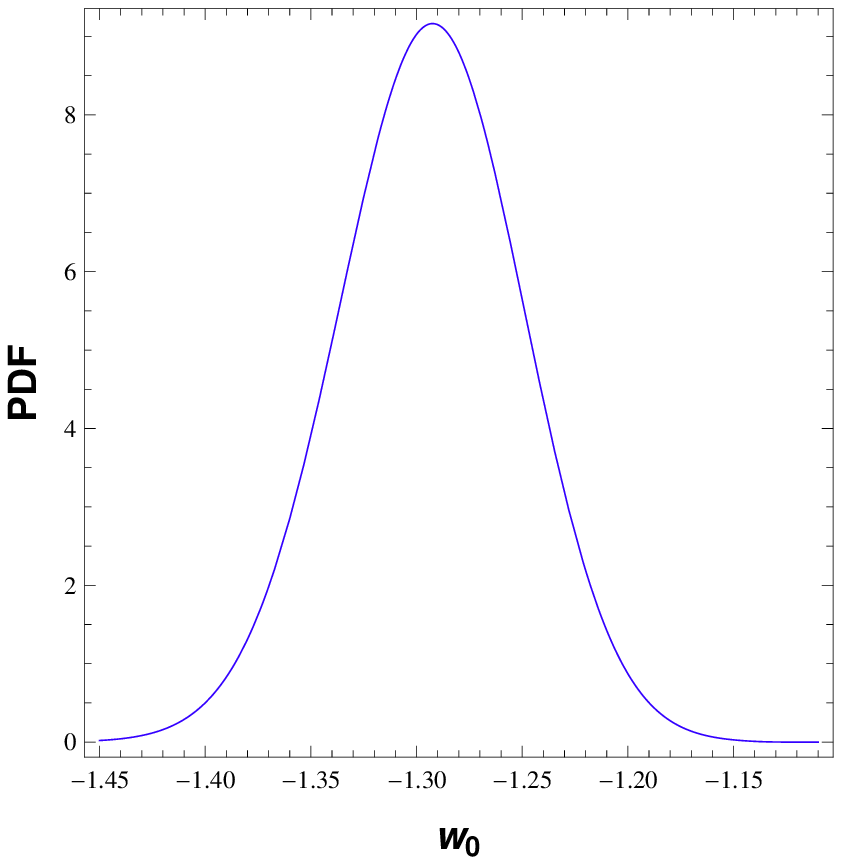}
\end{minipage} \hfill
\begin{minipage}[t]{0.3\linewidth}
\includegraphics[width=\linewidth]{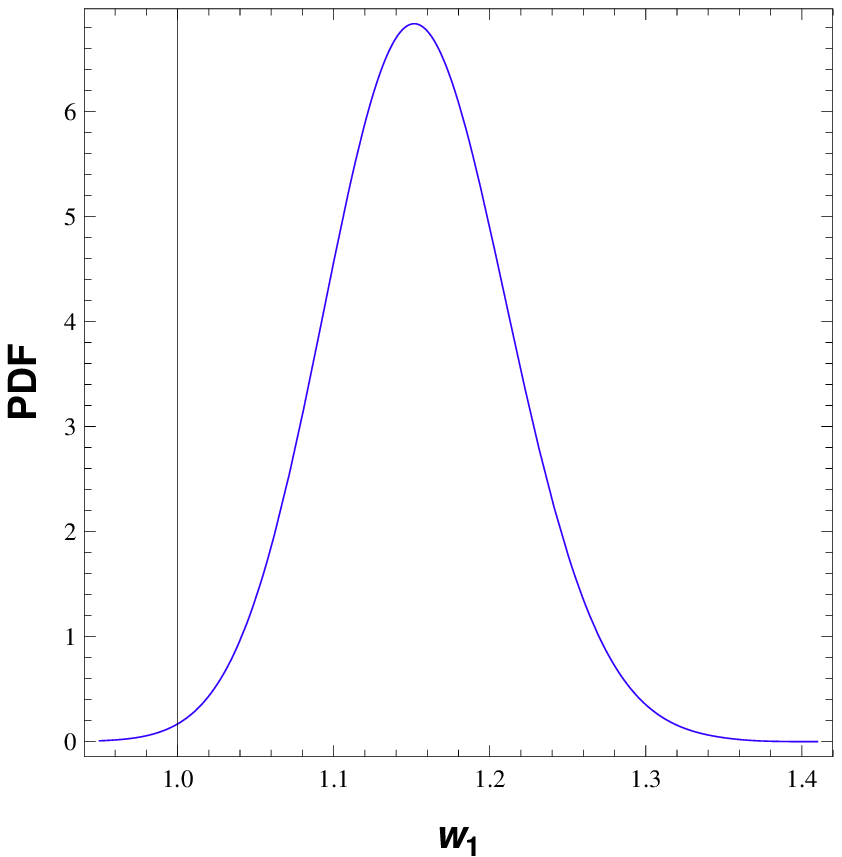}
\end{minipage} \hfill
\begin{minipage}[t]{0.3\linewidth}
\includegraphics[width=\linewidth]{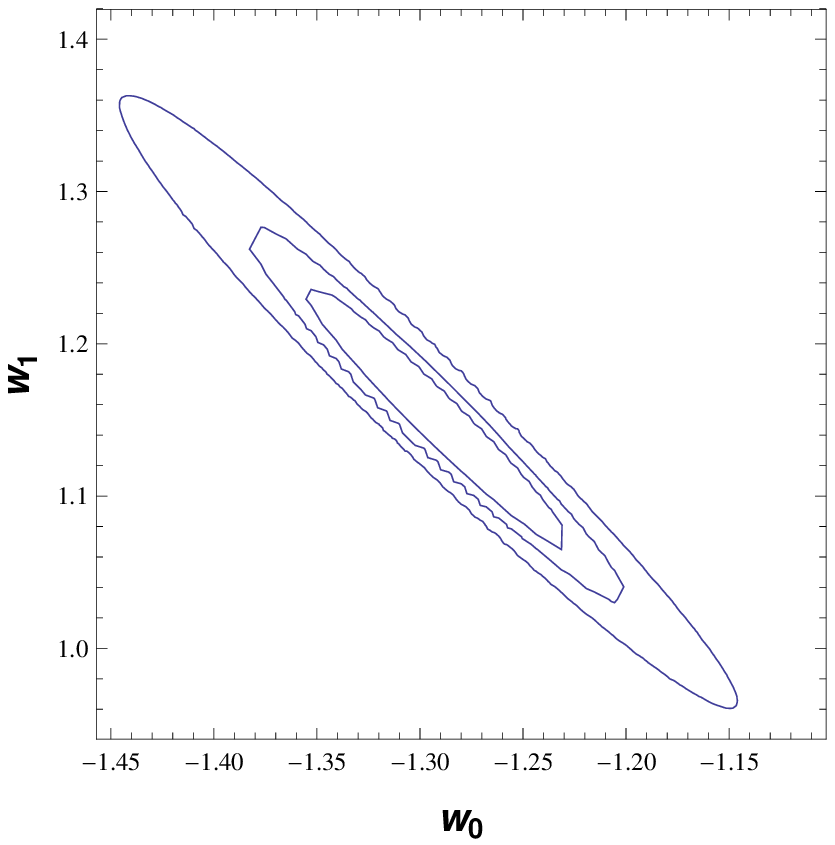}
\end{minipage} \hfill
\caption{{\protect\footnotesize Ricci-scale cutoff with CPL parametrization. Left panel: one-dimensional PDF for $w_0$. Center panel: one-dimensional PDF for $w_1$. Right panel: Contour plots for the $1\sigma$, $2\sigma$ and
$3\sigma$ confidence levels.
}}
\label{ricci}
\end{figure}
\end{center}

According to Eq.~(\ref{HRCPL}), there are four free parameters, $w_0$, $w_1$, $r_0$ and $H_0$. The minimum value of $\chi^2$ is
used to fix the values for $r_0$ and $H_0$, such that again two free parameters are left.
The estimations for the parameters $w_0$ and $w_1$ are $w_0 = - 1.29^{+0.08}_{-0.09}$ and $w_1 = 1.15^{+0.12}_{-0.11}$. Typically, we have also $0.3 < r_0 < 0.5$ according to the set of observational data used. The one- and two-dimensional PDFs are
displayed in figure \ref{ricci}.

\subsubsection{Interaction $Q = 3H\beta\rho_H$}
\label{ricciint}
\begin{center}
\begin{figure}[!t]
\begin{minipage}[t]{0.3\linewidth}
\includegraphics[width=\linewidth]{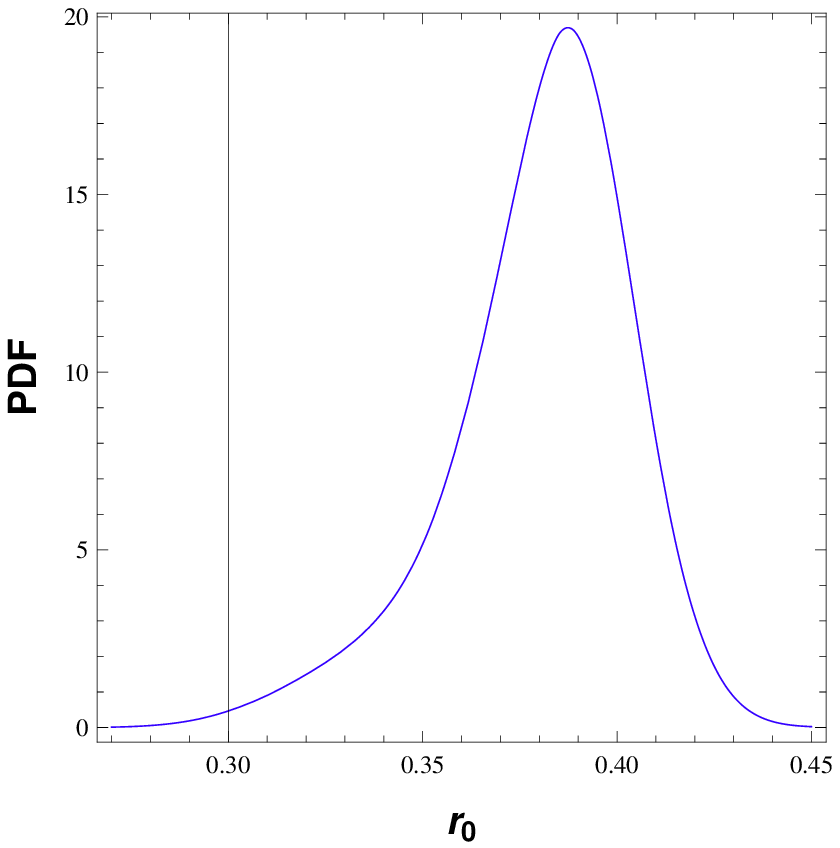}
\end{minipage} \hfill
\begin{minipage}[t]{0.3\linewidth}
\includegraphics[width=\linewidth]{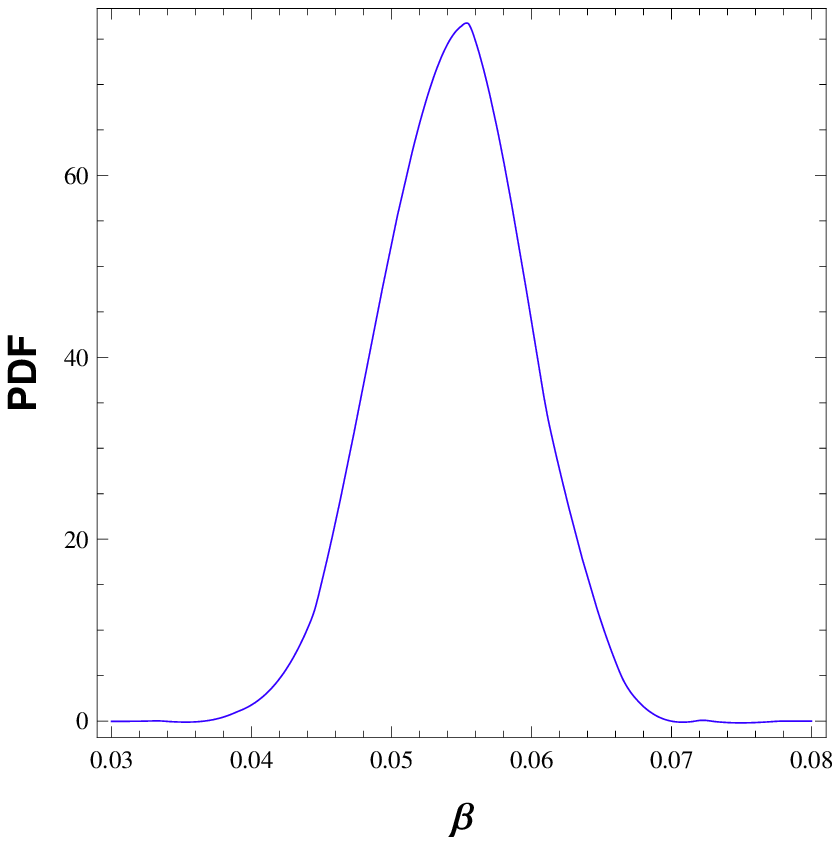}
\end{minipage} \hfill
\begin{minipage}[t]{0.3\linewidth}
\includegraphics[width=\linewidth]{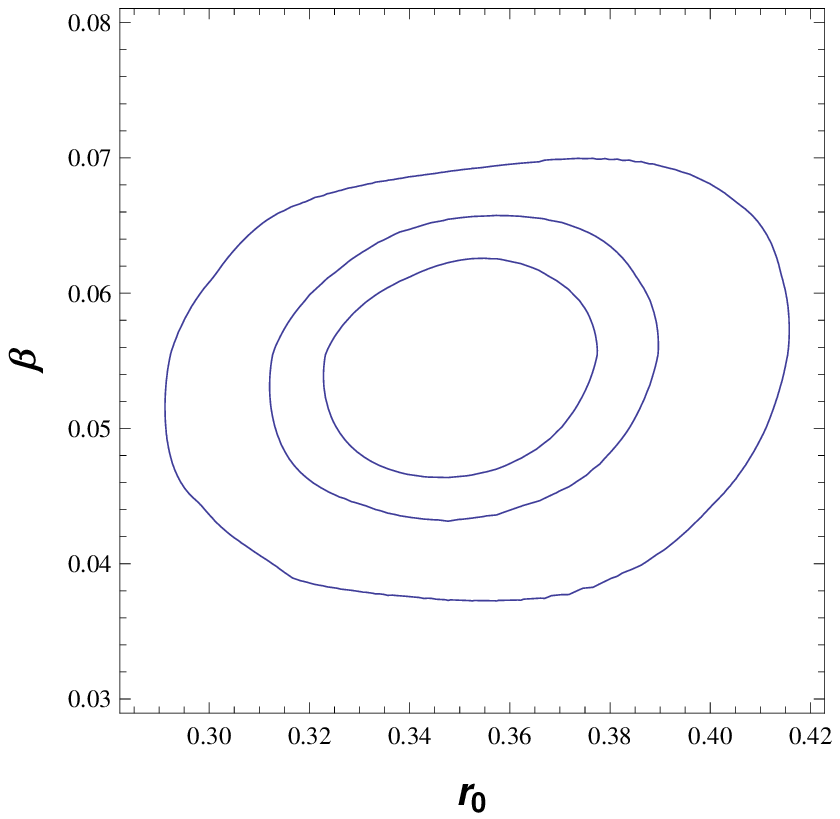}
\end{minipage} \hfill
\caption{{\protect\footnotesize Ricci-scale cutoff with interaction $Q = 3H\beta\rho_H$. Left panel: one-dimensional PDF for $r_0$. Center panel: one-dimensional PDF for $\beta$. Right panel: Contour plots for the $1\sigma$, $2\sigma$ and
$3\sigma$ confidence levels.
}}
\label{riccibis}
\end{figure}
\end{center}

In this
particular case the four free parameters are: $r_0$, $w$, $\beta$ and
$H_0$ (cf. Eq.~\ref{HRint}). Imposing $w = - 1$, and fixing $H_0$ as described above, we obtain the following estimation for the other two parameters at $2\sigma$: $r_0 = 0.35^{+0.04}_{-0.03}$ and
$\beta = 0.05^{+0.01}_{-0.01}$. The corresponding PDFs are shown in figure \ref{riccibis}.

\par
The different models may be properly compared among themselves and also with the $\Lambda$CDM reference model, using appropriate
statistical criteria. Two options are the Akaike Information Criterion ($AIC$), and the Bayesian Information Criterion ($BIC$), which allow us to compare models with different degrees of freedom.
The $AIC$ criterion uses the formula $AIC = \chi^2_{min} + 2\nu$ \cite{akaike}, where $\nu$ is the number of degrees of freedom; the
$BIC$ criterion \cite{schwarz} is based on the expression $BIC = \chi^2_{min} + 2\nu\ln N$, where $N$ is the number of observational points. The smaller the resulting numbers in both expressions, the higher the quality of the corresponding model.
The results are given in table \ref{comp}.
It is obvious that the $\Lambda$CDM model is the clear winner according to both of the criteria. The worst values are obtained for the Ricci-scale model with CPL parametrization and for the future-event-horizon model with $\xi = 1$. For the other models we find values that do not differ much from each other.

It is convenient to classify a model with respect to the differences $\Delta$AIC and $\Delta$BIC between its
$AIC$ and $BIC$ values, respectively, and the corresponding values for a reference model. This establishes a scale (Jeffreys' scale) which allows for a ranking of different models according to the magnitude of their differences  $\Delta$AIC and $\Delta BIC$ \cite{liddle}. The smaller the difference to the lowest $AIC$ or $BIC$ values, here those of the $\Lambda$CDM model, the better the model. For differences less than $2$, there is strong support for the model under consideration.  If $\Delta AIC (\Delta BIC) < 5$ the model is still weakly supported. Models with $\Delta AIC (\Delta BIC) > 10$ should be considered as strongly disfavored. By inspection, it follows from table \ref{comp}, that, using the $AIC$ criterion, the Hubble-scale models, and the interacting Ricci-scale model remain competitive with the $\Lambda$CDM model.
The other candidates are still weakly supported. Applying, however, the $BIC$ criterion, all the holographic models
appear to be ruled out.  This kind of contradiction in using
different evaluation criteria is well known in the literature, see, e.g., \cite{syz}.

It should be  mentioned, that the assessment of competing models also strongly depends on the choice of priors. Such a choice may radically change the final results from the $AIC$ or $BIC$ criteria.
For example, the seven years WMAP data indicate that $H_0 = 71.4$ for the $\Lambda$CDM model \cite{komatsu}. If this result is imposed as a prior for the holographic models, all estimations for the $AIC$ and $BIC$ values change and the overall picture is modified. As an example, let us consider the interacting Ricci-scale model (cf. Eqs.~(\ref{Hint}) and (\ref{HRint})).
It has three free parameters, $H_{0}$, $r_0$ and $\beta$. If we fix
$r_0 = 0.4$, two free parameters are left, and now this model has the same
number of degrees of freedom as the $\Lambda$CDM model. Under this condition we obtain $\chi^{2}_{total} = 550.99$,
essentially the same $\chi^{2}_{total}$ as for the $\Lambda$CDM model. Since the number of free parameters is the same, both models may be considered equally competitive. Hence, the imposing of priors may
change substantially the final result of the evaluation based on the $AIC$ and $BIC$ selection criteria.
Moreover, also priors of structure formation may alter the final evaluation.

\begin{table}[!t]
\begin{center}
\begin{tabular}{|c|c|c|c|c|c|c|c|}\hline
Model&FP(MP)&$SN(AIC)$&$SN(BIC)$&$H(AIC)$&$H(BIC)$&Total($AIC$)&Total($BIC$)\\ \hline
$\Lambda CDM$&$2(2)$&$546,70$&$567.99$&$12.07$&$18.33$&$554.78$&$576.16$\\ \hline
Hubble($\Omega_{m0}=0.25$)&$3(2)$&$548.59$&$580.53$&$14.02$&$23.41$&$556.62$&$588.69$\\ \hline
Hubble($\Omega_{m0}=0.30$)&$3(2)$&$548.59$&$580.53$&$14.02$&$23.41$&$556.62$&$588.69$\\ \hline
Event horizon($\xi = 1$)&$3(2)$&$550.49$&$582.43$&$14.53$&$23.92$&$559.03$&$591.10$\\ \hline
Event horizon($\xi = 2$)&$3(2)$&$548.79$&$580.72$&$14.09$&$23.48$&$556.88$&$588.95$\\ \hline
Event horizon($\xi = 3$)&$3(2)$&$548.95$&$580.89$&$14.19$&$23.58$&$557.15$&$589.22$\\ \hline
Ricci (CPL param.)&$4(2)$&$551.44$&$594.02$&$16.01$&$28.53$&$559.45$&$602.22$\\ \hline
Ricci($Q=3H\beta\rho_H$)&$3(2)$&$548.66$&$580.59$&$14.02$&$23.41$&$556.67$&$588.75$ \\ \hline
\end{tabular}
\end{center}
\caption{$AIC$ and $BIC$ criteria for the  assessment of the considered models. In the second column the number of free parameter of each
model is specified as well as the number of marginalized parameters (in parenthesis). The difference between the numbers of free and marginalized parameters leads to the number of fixed parameters.}  \label{comp}
\end{table}
\par
Another aspect that should be stressed is that $\chi^2_{min} \sim 550$ seems to be a saturation value for any
possible fitting using smooth functions. The reason is the dispersion in the SNIa and $H(z)$ data. The fact that this value is {\it grosso modo} achieved by the
$\Lambda$CDM model with just two free parameter seems to imply that any other model with more degrees of freedom would be automatically disfavored.
But, using priors and other observational tests, like those from a perturbative analysis, may
change the final conclusions again.

\section{Conclusions}
\label{conclusions}

Holographic DE models are based on a field theoretical relation between cutoffs in the ultraviolet and in the infrared energy regions. Different choices of the infrared cutoff give rise to different models.
We have presented here a detailed analysis of models with cutoffs at the Hubble scale, the future event horizon and a scale proportional to the Ricci length. Special emphasis was put on the role of interactions with a DM component and on the relation between the EoS parameter and the ratio of the energy densities of DM and DE.
We considered interaction-free limits of the dynamics and scaling solutions, i.e., solutions for which the energy density ratio of the dark components remains constant.
In order to constrain the parameter spaces for the general dynamics, a Bayesian statistical analysis has been performed, using data from supernovae type Ia, the history of the Hubble parameter, the position of the first acoustic peak in the CMB anisotropy spectrum and baryon acoustic oscillations.
In detail, our results can be summarized as follows. \\
(i) Hubble scale cutoff: for this choice a negative EoS parameter is a pure interaction effect \cite{DW,HDE}. The non-interacting limit cannot represent a DE component. Independently of the specific form of the interaction, the energy density ratio remains constant. There is no direct relation between this ratio and the effective EoS parameter.  For  suitable interactions, a transition from an early matter dominated epoch to an epoch of accelerated expansion can be obtained . We have assumed here an interaction rate proportional to a power of the Hubble rate. This gives rise to the dynamics of a generalized Chaplygin gas with the $\Lambda$CDM dynamics as a special case. The results of the statistical analysis are given in subsection \ref{hubbleanalysis}  and in figures \ref{hubble0.25} and \ref{hubble0.3}.\\
(ii) Future-event-horizon cutoff: in this case, there exist a direct relation between the energy density ratio and the EoS parameter. On the other hand, the effective EoS parameter, i.e., the parameter that includes the interaction, does not depend directly on the ``bare" EoS parameter. Assuming a power-law behavior of the energy density ratio, the cosmological dynamics was solved analytically for several values of the power. Using the preferred values of the statistical analysis in figures \ref{xi1}, \ref{xi2} and \ref{xi3} of subsection \ref{ehanalysis} as well as well as in table \ref{tab2}, we found three possibilities. In the first example, the present
effective EoS parameter has a value larger than $-1$. Also for $a \gg 1$ it will never cross the phantom divide.
The second example has a present value of the effective EoS larger than $-1$ as well, but this parameter will become smaller than $-1$ for large values of the scale factor. For the third example, both the present and the future values of the EoS parameter are of the phantom type.
Different from the Hubble-scale cutoff, there exists a non-interacting limit of the dynamics. Scaling solutions do exist as well. For all the three cutoff cases discussed here, these solutions can be mapped on an equivalent scalar-field dynamics with an exponential potential.
All models with a future-event-horizon cutoff suffer from the drawback that, although they may fit the data, they cannot reproduce an early matter dominated epoch.\\
(iii) Ricci-scale cutoff: here, there exists a linear relation between the EoS parameter and the energy-density ratio. The general dynamics was solved for the CPL parametrization. The results of the statistical analysis of subsubsection \ref{CPL} are displayed in figure \ref{ricci}.
Furthermore, we solved the dynamics for an interaction proportional to the DE density. The Bayesian analysis provided us with figure \ref{riccibis} of subsubsection \ref{ricciint}.
The noninteracting limit, which exists only for time-varying equations of state, has the interesting property that the energy-density ratio approaches a constant, finite value for $a \ll 1$ which is only about ten time larger than the present value. Nevertheless, the equation of state approaches that for dust, thus recovering an early matter dominated period.

Tables \ref{tab1} and \ref{comp} present  overviews of all the cases considered here.
The results of table \ref{comp} show, that the $\Lambda$CDM model remains the most favored option.
According to the $AIC$ criterion, the Hubble-scale models and the interacting Ricci-scale model can be considered as
competitive since they have $\Delta AIC <2$. The other models are still weakly supported. However, all of the models have $\Delta BIC >10$, which indicates that they are strongly disfavored according to the $BIC$ criterion.
The situation changes if certain priors are used, which reduce the number of free parameters to the number of free parameters of the $\Lambda$CDM model.
We demonstrated, that under these conditions the relevant $\chi^{2}$ values for the Ricci case, e.g.,  may be competitive with that of the $\Lambda$CDM model.
Our study has to be considered preliminary also in the sense that it is restricted to the homogeneous and isotropic background dynamics.
Nevertheless, we believe that a systematic investigation  as performed in this paper, may provide an idea about basic dynamical properties of the different models and subclasses of them.
What remains to be shown for a more advanced assessment is to consider the perturbation dynamics and to calculate
the matter power spectrum and the anisotropy spectrum of the CMB.
For a model to be competitive one has to require that the preferred parameter values of the perturbative analysis
coincide with those of the background dynamics. This issue will be the subject of a future investigation.

\acknowledgments{This work was supported by the ``Comisi\'{o}n
Nacional de Ciencias y Tecnolog\'{\i}a" (Chile) through the
FONDECYT Grant No. 1070306 (SdC) and  No. 1090613 (RH and SdC).
J.F. and W.Z. acknowledge support by ``FONDECYT-Concurso incentivo a la
Cooperaci\'{o}n Internacional" No. 1090613 and by CNPq (Brasil).}

\end{document}